\documentclass[prl,reprint,floatfix,twocolumn, nofootinbib, amsmath,amssymb]{revtex4-2}

\usepackage[final]{graphicx}
\usepackage{xcolor}
\usepackage{comment}
\usepackage{dcolumn}% Align table columns on decimal point
\usepackage{bm}% bold math
\usepackage{physics}
\usepackage{hyperref}% add hypertext capabilities
\hypersetup{
    colorlinks=true,
    urlcolor=blue,
    linkcolor=magenta, %sections, equatiions
    filecolor=magenta,      
    citecolor=cyan,
    pdftitle={Overleaf Example},
    pdfpagemode=FullScreen,
    }
\usepackage{tikz}
\usepackage{ulem}

\begin{document}

%\preprint{APS/123-QED}

\title{Testing the Wineland Criterion with Finite Statistics}
%\title{Has one surpassed the standard quantum limit \\ by attempting to squeeze spin systems?}
%\title{Have we gone below the standard quantum limit on spin {\color{red}{squeezed}} systems?}% Force line breaks with \\
\author{E. S. Carrera}
\thanks{These authors contributed equally to this work}

\author{Y. Zhang}
\thanks{These authors contributed equally to this work}
\author{J-D. Bancal}%
%\email{jbancal@ipht.fr}
\author{N. Sangouard}
%\email{nicolas.sangouard@ipht.fr}
\affiliation{%
 Université Paris-Saclay, CEA, CNRS, Institut de physique théorique, 91191, Gif-sur-Yvette, France}%


%\affiliation{
%\affiliation{
 %Third institution, the second for Charlie Author
%\affiliation{%


\date{\today}% It is always \today, today,

\begin{abstract}
The Wineland parameter aims at detecting metrologically useful entangled states, called spin-squeezed states, from expectations and variances of total angular momenta. {However, efficient strategies for estimating this parameter in practice have yet to be determined and in particular, the effects of a finite number of measurements remain insufficiently addressed. We formulate the detection of spin squeezing as a hypothesis-testing problem, where the null hypothesis assumes that the experimental data can be explained by non-spin-squeezed states. Within this framework, we derive upper and lower bounds on the p-value to quantify the statistical evidence against the null hypothesis.} By applying our statistical test to data obtained in multiple experiments, we are unable to reject the hypothesis that non-spin squeezed states were measured with a p-value of 5\% or less in most cases. We also find an explicit non-spin squeezed state according to the Wineland parameter reproducing most of the observed results with a p-value exceeding 5\%. More generally, our results provide a rigorous method to establish robust statistical
evidence of spin squeezing from the Wineland parameter in future experiments,
accounting for finite statistics.
\end{abstract}

%\begin{description}
%\end{description}
%\end{abstract}

\maketitle

\paragraph{Introduction ---}
Quantum metrology leverages quantum mechanical effects to surpass the limitations of classical measurement techniques~\cite{Giovannetti2004,Giovannetti2011}, notably the standard quantum limit. Among the quantum states that can be utilized to enhance measurement precision~\cite{Mitchell2004,Tomohisa2007,Pryde2017,ZhaoSiRan2021,Aasi2013,Guo2020,Tse2019,Xia2020}, spin-squeezed states are receiving significant attention~\cite{meyer2001,Bornet2023,Franke2023,Bohnet2016,Strobel2014,Riedel2010,Ockeloen2013,SchleierSmith2010,Leroux2010,LouchetChauvet2010,Bohnet2014,Sewell2012}. The condition for spin-squeezing is usually expressed in terms of the first and second moments of the angular momentum operator which is defined for $N$ distinguishable $1/2$ spins as 
\begin{equation}
J_{\alpha} = \frac{1}{2} \sum_{i=1}^{N} \sigma_{\alpha}^{(i)}
\end{equation}
where $\alpha=x,y,z$ specifies the direction and $\sigma_{\alpha}^{(i)}$ is the corresponding Pauli matrix for spin $i$. More precisely, a state that fulfills~\cite{Wineland1992}
\begin{equation}
\label{wineland}
\xi^2_W  = \frac{N \text{ Var}(J_{\Vec{n}_{\perp}})}{\langle J_{\vec{n}} \rangle ^2 } <1
\end{equation}
is said to be (metrologically) spin-squeezed along the $\vec{n}_\perp$ direction. Here, $\langle J_{\Vec{n}} \rangle$ denotes the expectation value of the spin operator along the direction $\vec{n}$ and Var$(J_{\Vec{n}_{\perp}})$ is the variance of the spin in a 
direction $\Vec{n}_{\perp}$ orthogonal to $\vec{n}$.
$\xi^2_W$ is called the Wineland parameter. 
Put simply, $\xi^2_W<1$ witnesses a class of quantum states where the collective spin variance is reduced in one direction while being increased in the orthogonal direction~\cite{Wineland1992,kitagawa1993}. In a Ramsey experiment, $\xi^2_W$ is established as the ratio of phase sensitivity of the state of interest $\hat{\rho}$ compared to a coherent spin state. Hence $\xi^2_W<1$ witnesses states having the potential to beat the standard quantum limit. The advantage of spin-squeezed states in metrology is due to quantum correlations between spins~\cite{kitagawa1993}. Specifically $\xi^2_W<1$  witnesses entanglement~\cite{Sorensen2001}. The ability to observe multiple spin states that are both metrologically useful and entangled through collective spin measurements has driven numerous experimental implementations involving up to almost a million spins~\cite{meyer2001,Bornet2023,Franke2023,Bohnet2016,Strobel2014,Riedel2010,Ockeloen2013,SchleierSmith2010,Leroux2010,LouchetChauvet2010,Bohnet2014,Sewell2012}. 

\smallskip
\noindent
{Experimentally confirming spin-squeezing has traditionally relied on the error bar associated to the estimation of the Wineland parameter.} {However, to demonstrate evidence of spin-squeezing, one should quantify the probability that the observed statistics, obtained from a limited number of experimental repetitions, could result from measurements on a state with no spin-squeezing, i.e. a state satisfying $\xi^2_W \geq 1$.} {Error bars only quantify the spread of the measurement outcomes. They do not provide information about the likelihood that the observed data could have been produced by non-spin-squeezed states~\cite{Krzywinski2013}.} {A more comprehensive statistical framework is thus needed for testing the Wineland criterion in practice. This challenge was highlighted in Refs.~\cite{Wagner17}, where a similar issue emerged in the context of Bell correlation witnesses in many-spin systems.}

\smallskip
\noindent
{
In this letter, we develop a rigorous statistical framework for establishing robust evidence of spin squeezing based on the Wineland parameter, explicitly accounting for finite-size effects. Our approach focuses on quantifying the probability (p-value) that the observed experimental data could arise from a non-spin-squeezed state, thereby formulating the problem as a hypothesis test.} {Unlike the approach in Ref.~\cite{jan2024}, our analysis does not test a specific non-spin-squeezed state and enables the rejection of the hypothesis that \textit{any} non-spin squeezed states could have produced the observed statistics with a specified p-value.} {
Our findings indicate that most existing experiments do not sufficiently account for finite-size effects, with only a limited number of cases capable of rejecting the hypothesis test. In many situations, an insufficient number of measurements has been performed which result in a high probability that a non-spin-squeezed state could mimic the experimental outcomes, particularly when the total spin number is large.}

\bigskip
\paragraph{Traditional approach ---} {The verification of spin squeezing in experiments is typically based on the statistical estimation of the Wineland parameter $ \xi^2_W $. Conventional approaches involve measuring the total spin $ \langle J_{\vec{n}} \rangle$ and the variance $ \text{Var}(J_{\vec{n}_{\perp}}) $, followed by an evaluation of the uncertainty in $ \xi^2_W $ from its error bar. A state is conventionally classified as spin-squeezed if the estimated value of $ \xi^2_W $ falls below 1 with a sufficiently small error bar. However, error bars do not constitute a rigorous test for statistical significance~\cite{Krzywinski2013}. In particular, they fail to account for the probability that statistical fluctuations may cause non-spin-squeezed states to satisfy $\xi^2_W < 1$, leading to potential misclassification. To illustrate this risk, we conduct a Monte Carlo simulation by repeatedly measuring a non-spin-squeezed state of $9$ spins with $\xi_{W}^2=1$, as detailed in {End Matter}. The results show that even with a large number of measurements resulting in a small error bar, statistical fluctuations can still cause a non-spin-squeezed state to be mistakenly classified as spin-squeezed.}

\bigskip
\paragraph{Methods ---} {
We here focus on null-hypothesis significance testing, with the p-value quantifying the probability of obtaining a Wineland parameter at least as extreme as the value actually observed, under the null hypothesis $H_0$ that the measured state is non-spin-squeezed.} {A small p-value thus provides strong evidence against $H_0$, suggesting that the observed data is unlikely under {the assumption that} they result from measurements of a non-spin-squeezed state. To ensure the p-value remains independent of any specific quantum state or statistical distribution, we determine an upper bound by considering all possible non-spin-squeezed states.}

% \smallskip
% \noindent
% \textcolor{blue}{Hypothesis testing evaluates spin squeezing by first defining a \textit{null hypothesis} $H_0$, which assumes the state is non-spin-squeezed. To assess this, a \textit{test statistic} is computed from the data, providing a quantitative measure to compare against $H_0$. We define:}
\smallskip
\noindent
{To formulate the null-hypothesis significance test, we introduce the normalized spin angular momentum operator $ Q_{(\cdot)} = 2J_{(\cdot)}/N $, define the following parameter}
\begin{equation}
\label{eq3}
\Gamma = N \text{Var} (Q_{{\vec{n}_\perp}}) - \langle Q_{\vec{n}} \rangle^2.
\end{equation}
{and use the sample variance and sample mean as unbiased estimators for $\text{Var}(Q_{\vec{n}_{\perp}})$ and $\langle Q_{\vec{n}}\rangle^2$. In the asymptotic limit (infinite measurement repetitions), a spin system is considered spin-squeezed if the corresponding state satisfies $\Gamma < 0$.} {While Eq.~\eqref{eq3} provides a natural alternative criterion for spin squeezing, it involves nonlinear terms. Existing concentration inequalities struggle to effectively account for these nonlinearities, resulting in an overly conservative upper bound on the p-value.  As a result, the predicted number of measurements required to achieve a small p-value is overestimated (see Supplemental Material, Secs.~II and V).} 

\smallskip
\noindent
{To address these challenges, we propose a family of parameters
\begin{equation}
\Gamma_{c}=
N\langle Q^{2}_{\vec{n}_{\perp}}\rangle - f_{\alpha,\beta}(\langle Q_{\vec{n}_{\perp}}\rangle,\langle Q_{\vec{n}}\rangle),
\label{eq4}
\end{equation}
where each element is defined by a fixed value of $c=(\alpha,\beta)$. The function $f_{\alpha, \beta}(x, y)$ is the tangent plan to $Nx^2+y^2$ at the point $ c=(\alpha,\beta)\in [-1,1] $
\begin{equation}
    f_{\alpha, \beta}(x, y) =  2\alpha Nx + 2\beta y - N\alpha^2 - \beta^2. 
\end{equation}
Importantly, $f_{\alpha, \beta}(x, y)$ serves as an affine lower bound for $ Nx^2+y^2 $ for any $(x,y)$ and $(\alpha,\beta)$. This ensures that $\Gamma_{c}\geq N\langle Q^{2}_{\vec{n}_{\perp}}\rangle - N\langle Q_{\vec{n}_{\perp}}\rangle^2 - \langle Q_{\vec{n}}\rangle^2 = \Gamma $ for any choice of  $c=(\alpha,\beta)$. As a result, quantum states satisfying $\Gamma_{c}<0$ can be classified as spin-squeezed.}

\begin{figure*}[ht]
    \centering
    \begin{minipage}{\textwidth} % Adjusting the width to full text width
        \centering
        \begin{tikzpicture}
            \node[anchor=south west,inner sep=0] (image) at (0,0) {\includegraphics[width=1\textwidth]{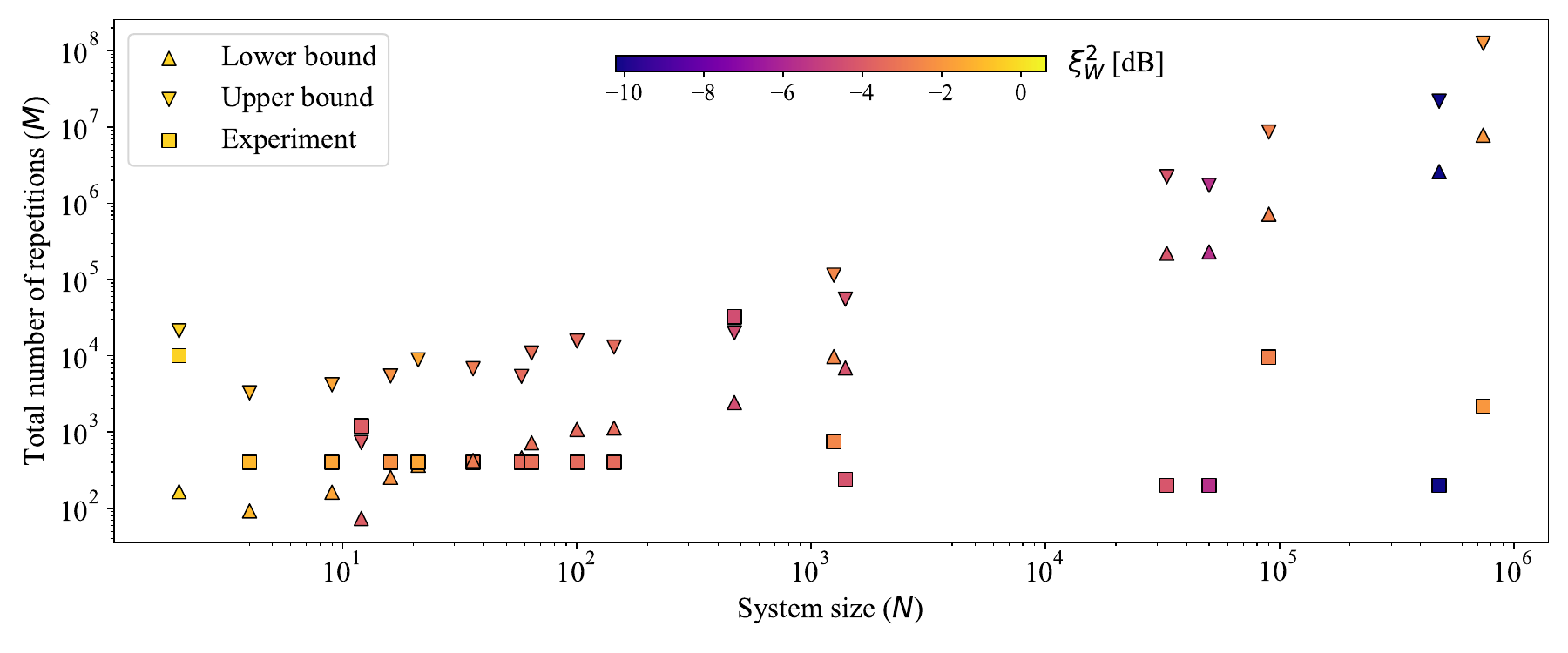}};
            \begin{scope}[x={(image.south east)},y={(image.north west)}]
                % Citations inside the plot
                \node[right] at (0.094,0.42) {\cite{meyer2001}};
                \node[right] at (0.138,0.26) {\cite{Bornet2023}};
                \node[right] at (0.19,0.327) {\cite{Bornet2023}};
                \node[right] at (0.21,0.384) {\cite{Franke2023}};
                \node[right] at (0.228,0.327) {\cite{Bornet2023}};
                \node[right] at (0.247,0.25) {\cite{Bohnet2016}};
                \node[right] at (0.28,0.33) {\cite{Bornet2023}};
                \node[right] at (0.303,0.25) {\cite{Bohnet2016}};
                \node[right] at (0.325,0.35) {\cite{Bornet2023}};
                \node[right] at (0.348,0.25) {\cite{Bornet2023}};
                \node[right] at (0.375,0.25) {\cite{Bohnet2016}};
                \node[right] at (0.446,0.55) {\cite{Strobel2014}};
                \node[right] at (0.51,0.362) {\cite{Riedel2010}};
                \node[right] at (0.518,0.23) {\cite{Ockeloen2013}};
                \node[right] at (0.724,0.223) {\cite{SchleierSmith2010}};
                \node[right] at (0.75,0.296) {\cite{Leroux2010}};
                \node[right] at (0.788,0.42) {\cite{LouchetChauvet2010}};
                \node[right] at (0.897,0.225) {\cite{Bohnet2014}};
                \node[right] at (0.925,0.348) {\cite{Sewell2012}};
            \end{scope}
        \end{tikzpicture}
    \end{minipage}
    % References below the plot in two columns using minipage
    \vspace{1em} % Some space before the references
    \begin{minipage}{0.32\textwidth}
        \begin{itemize}
            \item \cite{meyer2001} Meyer et al. (2001)
            \item \cite{Bornet2023} Bornet et al. (2023)
            \item \cite{Franke2023} Franke et al. (2023)
            \item \cite{Bohnet2016} Bohnet et al. (2016)
        \end{itemize}
    \end{minipage}
    \begin{minipage}{0.32\textwidth}
        \begin{itemize}
            
            \item \cite{Strobel2014} Strobel et al. (2014)
            \item \cite{Riedel2010} Riedel et al. (2010)
            \item \cite{Ockeloen2013} Ockeloen et al. (2013)
            \item \cite{SchleierSmith2010} Schleier-Smith et al. (2010)
        \end{itemize}
    \end{minipage}
    \begin{minipage}{0.32\textwidth}
        \begin{itemize}

            \item \cite{Leroux2010} Leroux et al. (2010)
            \item \cite{LouchetChauvet2010} Louchet-Chauvet et al. (2010)
            \item \cite{Bohnet2014} Bohnet et al. (2014)
            \item \cite{Sewell2012} Sewell et al. (2012)
        \end{itemize}
    \end{minipage}
 % Caption for the figure
    \caption{
    %Comparison between the number of experimental realizations suggested by our bounds and the literature.} Here, we fix the p-value upper bound and lower bound at $p_{\text{opt}} = p^* = 0.05$. For the upper bound, we express the total  number of repetitions $M$ as a function of $\mathbb{E}(q_{\vec{n}_{\perp}})$, $\mathbb{E}(q_{\vec{n}})$, $\alpha$ and $\beta$. Since in some of the cited papers $\mathbb{E}(q_{\vec{n}_{\perp}})$ is not given, we consider $\mathbb{E}(q_{\vec{n}_{\perp}}) = 0$, $\mathbb{E}(q_{\vec{n}_{\perp}}^2) = \text{Var}_{\text{exp}}(q_{\vec{n}_{\perp}})$, and $\mathbb{E}(q_{\vec{n}}) = \mathbb{E}_{\text{exp}}(q_{\vec{n}})$. Then, we find the minimum number of repetitions by tuning $\alpha$ and $\beta$. For the lower bound, we set $\xi_W^2(\ket{\chi}) = \xi_{W\text{exp}}^2$. Then, we find the number of measurements from $M = \log(p^*) / \log(r)$. The subscript $exp$ indicates experimental information.
    Experimental achievements together with their statistical analysis. The experiments that have been reported in the litterature are identified by a square whose position specifies the total number of experimental repetitions ($M$) that has been realized and the system size ($N$). The color of the square gives the observed value of the Wineland parameter. Every square is accompanied by a number used to identify the reference as specified below the graph. The significance of the experimental outcomes are quantified by the number of measurements that are necessary and sufficient to reach a p-value of 5\%. Namely, the upper triangle (with the vertex pointing down) gives the total number of measurement repetitions that is sufficient to reject the hypothesis that non-spin squeezed states were measured. The lower triangle (with the vertex pointing up) specifies the number of measurements that is required in order to ensure that  a specific non-spin squeezed state (according to the Wineland parameter) cannot reproduce the observed statistics.
    }
     \label{plot_measures_lower_upperbounds_with_data}
\end{figure*}

\smallskip
\noindent
{To estimate $\Gamma_{c}$ in practice, we consider $M/2$ independent and identically distributed (i.i.d.) repeated processes ($M$ is the total number of experimental repetitions). In each processes $i$, the system undergoes two independent measurements. In the first measurement, the quantum state is prepared, and a measurement of  $Q_{\vec{n}}$ is performed, yielding the outcome $q^{(i)}_{\vec{n}} \in [-1,1]$. In the second measurement, the quantum state is re-prepared, and a measurement of  $Q_{\vec{n}_{\perp}}$ is performed,  obtaining the outcome $q^{(i)}_{\vec{n}_{\perp}} \in [-1,1]$. {Using these measurement outcomes, we define the  estimator of $\Gamma_c$  for each experimental processes $i$ as the random variable
\begin{equation}
\label{gammaci}
\widetilde{\Gamma}_{c,i}= N \left(q^{(i)}_{\vec{n}_{\perp}}\right)^{2} - f_{\alpha,\beta}\left( q^{(i)}_{\vec{n}_{\perp}}, q^{(i)}_{\vec{n}}\right).
\end{equation}
Given the $M/2$ repetitions, $\Gamma_c$ can be estimated by using the following statistical estimator
\begin{equation}
\label{gammac}
\widetilde{\Gamma}_{c} =\frac{2}{M}\sum_{i=1}^{M/2} \widetilde{\Gamma}_{c,i}.
\end{equation}
Let $\gamma_{c} < 0$ be the observed value of $\widetilde{\Gamma}_{c}$ from a given experimental realization.} We are interested in the  p-value, i.e. the probability that the estimator $\widetilde{\Gamma}_{c}$ takes a value equal to or smaller than $\gamma_{c}$, assuming the measurements are performed on non-spin-squeezed states (the null hypothesis $H_0$). We name it explicitly as $P(\widetilde{\Gamma}_{c} \leq \gamma_{c} \, | \, H_{0})$. Importantly, $P(\widetilde{\Gamma}_{c} \leq \gamma_{c} \, | \, H_{0})$ must be estimated by considering all possible non-spin-squeezed states and all possible statistical distributions. If $P(\widetilde{\Gamma}_{c} \leq \gamma_{c} \, | \, H_{0})$ is less than a predetermined significance level, such as 0.05, we conclude that the system of interest exhibits spin squeezing.}

\bigskip
\paragraph{Establishing upper bounds on the p-value ---} {Concentration inequalities~\cite{McDiarmid1989,McDiarmid1998} can be used to upper bound  $P(\widetilde{\Gamma}_{c} \leq \gamma_{c} \, | \, H_{0})$. As shown in Supplemental Material, Sec. III, the Bernstein inequality for example gives} 
{
\begin{equation}
\label{bers_edi_2}
P(\widetilde{\Gamma}_{c}\leq \gamma_{c} | H_0)\leq \exp{\dfrac{\gamma^{2}_{c} M/2}{2\Gamma^{c}_{1}\Gamma^{c}_{0}+2(\Gamma^{c}_{1}-\Gamma^{c}_{0})\gamma_{c}/3}}
\end{equation}
where
\begin{align}
\nonumber
\Gamma^{c}_{1}&=\max_{q^{(i)}_{\vec{n}_{\perp}}, q^{(i)}_{\vec{n}}} \widetilde{\Gamma}_{c,i}=N(1+|\alpha|)^2+(1+|\beta|)^2-1, \\ \nonumber
\Gamma^{c}_{0}&=\min_{q^{(i)}_{\vec{n}_{\perp}}, q^{(i)}_{\vec{n}}} \widetilde{\Gamma}_{c,i}=(1-|\beta|)^2-1.
\end{align}}
{The right-hand side of Ineq.~\eqref{bers_edi_2} is solely a function of the parameter $c = (\alpha, \beta)$ and the experimental observation $\gamma_c$. It can be minimized over $c = (\alpha, \beta)$ under the constraint that $\gamma_c<0$ in order to get the tightest upper bound on the p-value. This defines
$p_{\text{opt}}$ as
\begin{equation}
\label{popt}
p_{\text{opt}}= \inf_{\alpha, \beta \text{ s.t.}\gamma_c<0} \exp{\dfrac{\gamma^{2}_{c} M/2}{2\Gamma^{c}_{1}\Gamma^{c}_{0}+2(\Gamma^{c}_{1}-\Gamma^{c}_{0})\gamma_{c}/3}}.
\end{equation}
}

\bigskip
\paragraph{Guidelines for robust statistical evidence of spin squeezing in future experiments ---} {We now have all the ingredients to propose a concrete recipe that can be used to establish a robust statistical evidence of spin squeezing in future experiments. First, we consider $M/2$ measurements of both $Q_{\vec{n}}$ and $Q_{\vec{n}_{\perp}}$. This yields outcomes $q^{(i)}_{\vec{n}}$ and  $q^{(i)}_{\vec{n}_{\perp}}$ for $i\in [1,M/2]$. $\gamma_c$ is then deduced from the estimator $\widetilde{\Gamma}_{c}$ given in Eq.~\eqref{gammaci}. An upper bound on the p-value is obtained from the minimization given in Eq.~\eqref{popt} over $\alpha$ and $\beta$ such that $\gamma_c<0$. A small $p_{\text{opt}}$ provides strong
evidence against $H_0$, suggesting that the observed data is unlikely under {the assumption that} they result from measurements of a non-spin-squeezed state.}

\bigskip
\paragraph{Analysis of existing experimental realizations ---} 
To estimate what could have been achieved in previously implemented experimental realizations, we calculated the value of $\gamma_c$ from Refs.~\cite{meyer2001,Strobel2014, Riedel2010, Ockeloen2013,Bohnet2016, SchleierSmith2010, Leroux2010, LouchetChauvet2010, Bohnet2014, Sewell2012,Bornet2023,Franke2023}
by assuming that $\sum_{i=1}^{M/2} q^{(i)}_{\vec{n}_{\perp}}=0$. We increase $M$, optimizing $(\alpha, \beta)$, until the upper bound on the p-value achieves $5$\%. 
%We then vary the number of experimental repetitions $M$ for fixed values of $(\alpha, \beta)$ until the upper bound on the p-value achieves $5$\%. We then explore different values for the parameters $(\alpha, \beta)$ to minimize $M$. 
We then execute numerical simulations to show that the assumption $\sum_{i=1}^{M/2} q^{(i)}_{\vec{n}_{\perp}}=0$ is conservative, i.e.~it minimizes the number of experimental repetitions that are needed to achieve a given upper bound on the $p$-value, see Supplemental Material Sec.~IV. The value of $M$ that we find is reported in Fig.~\ref{plot_measures_lower_upperbounds_with_data} (triangle with the vertex pointing down) as a function of the system size for each experiment separately. The actual number of experimental repetitions conducted is also indicated (square). 
Except for the experiment involving 16 spins\footnote{In Ref.~\cite{Franke2023}, the authors specify that each measurement point reported in their paper is an average of 50 to 600 repetitions of experimental realizations. %To compute the proposed upper bound, we assumed that 600 realizations were performed in each directions, $\vec{n}$ and $\vec{n}_{\perp}$. 
The position of the corresponding square in Fig.~\ref{plot_measures_lower_upperbounds_with_data} assumes that 1200 repetitions in total were performed, as two measurements directions are involved . Actually, 728 realizations are enough to achieve an
upper bound on the p-value of 5\%.} reported in Ref.~\cite{Franke2023} and 470 spins\footnote{In Ref.~\cite{Strobel2014}, a tomography was performed using a total of 32500 experimental realizations.
%To compute the proposed upper bound, we assumed that 32500 measurements are performed in the each directions. 
The position of the corresponding square in Fig.~\ref{plot_measures_lower_upperbounds_with_data} hence assumes that 32500 realizations were performed.
Actually, 19970 realizations (9985 in each measurement direction) ensure a significance level of 5\%.} in Ref.~\cite{Strobel2014}, we cannot reject the hypothesis that non-spin squeezed states were measured with a p-value
less than or equal to 5\%. For most experiments, 
the discrepancy between the actual number of measurements and the number required to achieve an upper bound on the p-value of 5\% is at least a factor of 10, and can reach up to $10^{4}$ at the worst case. 

% \textcolor{red}{ For large-spin systems, this discrepancy grows exponentially, emphasizing the need for designs that balance spin numbers with measurement repetitions.}

\bigskip
\paragraph{Establishing lower bounds on the p-value ---} 
The upper bound on the p-value we derived before suggests that the number of experimental repetitions performed may be insufficient to rule out the possibility that non-spin-squeezed states were actually produced in most existing experimental realisations. To substantiate this claim, we follow the approach of Refs.~\cite{Raman2016,jan2024} and in particular, we consider the following state
\begin{equation}
    \label{lower state}
    \hat{\rho} = r \ket{\chi}\bra{\chi} + \dfrac{1-r}{2}\left( \ket{\pmb{\uparrow_{\Vec{n}_{{\perp}}}}}\bra{\pmb{\uparrow_{\Vec{n}_{{\perp}}}}} + \ket{\pmb{\downarrow_{\Vec{n}_{{\perp}}}}}\bra{\pmb{\downarrow_{\Vec{n}_{{\perp}}}}}\right)
\end{equation}
where $\ket{\chi}$ is a squeezed state with mean spin direction $\Vec{n}$, $\ket{\pmb{\uparrow_{\Vec{n}_{{\perp}}}}}=\ket{\uparrow_{\Vec{n}_{{\perp}}}}^{\otimes N} $, $\ket{\pmb{\downarrow_{\Vec{n}_{{\perp}}}}}=\ket{\downarrow_{\Vec{n}_{{\perp}}}}^{\otimes N}$ and $0\leq r\leq 1$. The state $\hat \rho$ is not considered as being spin squeezed by the Wineland parameter ($\xi_W^2 (\hat{\rho}) \geq 1$) for any values of $r$ satisfying
\begin{equation}
\label{rmax}
r \leq r_{\text{max}}= \dfrac{1}{2} \left( \kappa + \sqrt{\kappa^2 +\dfrac{4N}{\langle Q_{\Vec{n}}\rangle_{\ket{\chi}}^2} } \right)
\end{equation}
where $\kappa=\xi_W^2 \left(\ket{\chi} \right) - N /\langle Q_{\Vec{n}}\rangle_{\ket{\chi}}^2 $ (see Supplemental Material Sec.~VI).
When performing $M$ measurements on $\hat\rho$ with $r\leq r_{\text{max}}$, the probability to only sample the squeezed state $\ket{\chi}$ occurs with a probability $p=r^{M}\leq r_{\text{max}}^{M}=p^{*}$. Hence, $p^{*}$ establishes a lower bound on the p-value of a statistical test designed to reject the hypothesis that the measured state is of the form given in Eq.~\eqref{lower state}~\cite{Raman2016,jan2024}.

\smallskip
\noindent
For each experiment reported in Refs.~\cite{meyer2001,Strobel2014, Riedel2010, Ockeloen2013,Bohnet2016, SchleierSmith2010, Leroux2010, LouchetChauvet2010, Bohnet2014, Sewell2012,Bornet2023,Franke2023}, we compute the minimum number of measurements required to ensure that the observed statistics could not be reproduced by a non-spin-squeezed state of the form given by $\hat \rho$ with a probability $p^*$ sets to 5\%. This is derived from $\log(0.05)/\log(r_{\text{max}})$ where $\xi_W^2(\ket{\chi})$ and $\langle Q_{\Vec{n}}\rangle_{\ket{\chi}}^2$ are fixed to the observed values. The results, reported in Fig.~\ref{plot_measures_lower_upperbounds_with_data}, show that for most of the experiments involving more than 100 spins, the proposed non-spin squeezed state reproduces the observed results with a probability exceeding 5\%.

\bigskip
\paragraph{Discussion and Conclusion ---} {In this paper, we developed a rigorous analytical framework for detecting spin-squeezed states, with a particular focus on the impact of finite measurement repetitions on the detection results. Specifically, by linearizing $\xi_{W}^{2}$, we proposed a family of criteria for identifying spin-squeezed states. Using a hypothesis-testing approach, we derived upper bounds on the p-value, which can be used to provide a reliable statistical evidence of spin squeezing.}

\smallskip
\noindent
{The effects of finite measurement repetitions have long been overlooked in spin-squeezing detection experiments.  Our hypothesis-testing framework reveals that most existing experiments fail to pass the statistical test due to an insufficient number of measurements. This issue becomes particularly severe when the spin number is large, as the required number of measurements needed to reliably reject the null hypothesis increases significantly, often exceeding the practical capabilities of current experimental setups. However, this does not necessarily mean that spin squeezing is absent, but rather that the available data does not provide strong enough statistical support for claiming spin squeezing is produced. }

%Our findings suggest that future experimental designs should strike a balance between enhancing the degree of spin squeezing and increasing the total number of spins to optimize detection reliability and statistical robustness.} 

% We designed a linear estimator associated to the Wineland parameter and use the Bernstein inequality to quantify the significance of experimental data. This defines a proper recipe for estimating the Wineland parameter in practice. The first step consists in extracting the value $\gamma_c$ that the estimator $\tilde{\Gamma}_c$ takes on the observed statistics using Eq.~\eqref{gammaci} and then deducing the upper bound on the p-value from the minimization over $\alpha$ and $\beta$ of the right hand side of Eq.~\eqref{bers_edi_2} while  fixing the number of experimental repetitions $M$. %and optimizing over $\alpha$ and $\beta$.
% The null hypothesis associated to ``the state of interest satisfies $\Gamma\geq0$" can then be rejected if the p-value is below 0.05 or any other desired value.

\smallskip
\noindent
Our analysis provides an estimate of the number of measurements which is necessary (but not sufficient) to reject the null hypothesis. From a specific state, we showed that the p-value is lower bounded by $p^*=r_{\max}^M$ where $r_{\max}$ is specified from the expected values of $\xi_W^2$ and $\langle Q_{\Vec{n}}\rangle$, see Eq.~\eqref{rmax}. Note that $r_{\max}\geq (\sqrt{N^{2}+4N}-N)/2$ and hence, a number of measurements $M\geq \ln{0.05}/{\ln{[(\sqrt{N^{2}+4N}-N)/2]}}\sim -N\log(0.05)$ is necessary to achieve a p-value smaller than 5\%.

\smallskip
\noindent
Note that the p-value bounds we proposed in Eq.~\eqref{popt} could likely be improved. For instance, using more refined estimators may yield tighter bounds. Additionally, various concentration inequalities could be explored, and although several have already been tested (see Supplemental Material Sec. V), other choices might offer even stronger bounds. Similarly, while we have tested multiple states to construct the lower bound on the p-value, we cannot rule out the possibility that better states exist for this purpose. {Moreover, our statistical method could be extended to quantify finite-size effects on other squeezing or entanglement criteria~\cite{atomicensembles2018}.} We leave these extensions for future work. Nonetheless, our results already offer a solid framework for predicting and accurately assessing the outcomes of future experiments focusing on spin squeezing.

\bigskip
\paragraph{Acknowledgments ---} 
We thank Gregoire Misguich for helpful discussions. We acknowledge funding by the European High-Performance Computing Joint Undertaking (JU) under grant agreement No 101018180 and project name HPCQS and by a French national quantum initiative managed by Agence Nationale de la Recherche in the framework of France 2030 with the reference ANR-22-PETQ-0007 and project name EPiQ.

%\bibliography{apssamp2}% 


\newpage
\appendix*
{{\section{End Matter}
In experimental estimations of the Wineland parameter, standard error bars reflect only the statistical variability of the measured data and do not rigorously quantify the likelihood that the observed statistics could originate from non-spin-squeezed states. To highlight this limitation, we performed Monte Carlo simulations on a specific non-spin-squeezed state, demonstrating explicitly that small error bars do not necessarily correspond to small p-values. This reveals the potential risk of misclassifying non-spin-squeezed states as spin-squeezed when relying solely on error bar.
}}

\smallskip
\noindent
{{Specifically, we consider quantum states of the form given by Eq.~\eqref{lower state}
\begin{equation}
\hat{\rho} = r \ket{\chi}\bra{\chi} + \frac{1-r}{2}\left(\ket{\pmb{\uparrow_{\Vec{n}_{{\perp}}}}}\bra{\pmb{\uparrow_{\Vec{n}_{{\perp}}}}} + \ket{\pmb{\downarrow_{\Vec{n}_{{\perp}}}}}\bra{\pmb{\downarrow_{\Vec{n}_{{\perp}}}}}\right)
\end{equation}
where $|\chi\rangle$ is generated via an appropriate time evolution under a dipolar XY Hamiltonian (see Supplemental Material Sec.~I). In our simulations, we use a square lattice of $N=9$ spins and obtain a spin-squeezed state characterized by $\xi_W^2(|\chi\rangle)\approx 0.41036$, a mean spin polarization $\langle 2J_y/N\rangle_{\chi}\approx0.80812$, and a variance $\text{Var}(2J_x/N)_{\chi}\approx0.02977$. From the value of $\xi_W^2(|\chi\rangle)$, we construct a non-spin-squeezed state $\hat{\rho}$ satisfying $\xi_W^2(\hat{\rho})=1$, corresponding to parameters $r\approx0.96154$, $\langle2J_y/N\rangle_{\rho}\approx0.77704$ and $\text{Var}\left(2 J_{x}/N\right)_{\rho} \approx 0.06709$.

\smallskip
\noindent
The statistics that would be obtained from such a state are generated using Monte Carlo sampling.
Two data sets, each of size $M/2$, are produced ($M$ is the total number of samples). The first data set, $Q_{y}=\{y_1, \dots, y_{M/2}\}$, is sampled in the eigenbasis of $2J_{y}/N$. The second data set, $Q_{x}=\{x_1, \dots, x_{M/2}\}$, is sampled in the eigenbasis of $2J_{x}/N$. To establish
a comprehensive error estimation framework, the measurement outcomes $x_i$ and $y_j$ are assumed to be independent and identically distributed (i.i.d.). Their (unknown) true means are labeled ($\mu_x$, $\mu_y$) and true variances by ($\sigma_x^2$, $\sigma_y^2$).

\smallskip
\noindent
Recalling the definition of the Wineland parameter from Eq.~\eqref{wineland}, it can be conveniently rewritten as
\begin{equation}
\xi_W^2 = \frac{N\left(\mu_{x^2} - \mu_x^2\right)}{\mu_y^2},
\label{eq:xi_def}
\end{equation}
where $\mu_{x^2}$ is the true mean of the second-order moment of $x_i$ which satisfies $\mu_{x^2}= \sigma_{x}^2+\mu_{x}^2$. A natural approach to estimate $\xi_{W}^{2}$ in practice is to substitute the true expectations by the sample means. However, this introduces a systematic bias due to the nonlinear dependence on $\mu_x$ and $\mu_y$~\cite{becca2017quantum}.
To mitigate this issue, the jackknife method can be used~\cite{Franke2023,Strobel2014}.
Specifically, the bias-corrected jackknife estimation of $\xi_{W}^{2}$ is given by (see Supplemental Material Sec~I)
\begin{align}
\frac{M}{2}\xi_W^2(\bar{x}, \overline{x^2}, \bar{y}) - \left(\frac{M}{2} - 1\right)\overline{[\xi_W^2]},
\label{eq:error_bar}
\end{align}
where $\overline{x}=\frac{1}{M/2}\sum_{i=1}^{M/2}x_{i}$, $\overline{x^{2}}=\frac{1}{M/2}\sum_{i=1}^{M/2}x_{i}^{2},$ $\overline{y}=\frac{1}{M/2}\sum_{i=1}^{M/2}y_{i}$ are the sample means of datasets. \(\overline{[\xi_W^2]}\) is the mean of jackknife replicates, defined as $\overline{[\xi_{W}^2]}=\frac{1}{M/2}\sum_{i=1}^{M/2}\xi_{W}^2\left([x]_i,[x^2]_i,[y]_i\right)$ with $[x]_i = \frac{1}{M/2-1}\sum_{j\neq i}x_j$, $[y]_i = \frac{1}{M/2-1}\sum_{j\neq i}y_j$ the jackknife leave-one-out averages. %The bias of this jackknife estimator is of order $O(2/M)$. 

\smallskip
\noindent
The uncertainty in the jackknife estimation is quantified by the error bar
\begin{equation}
\sqrt{M/2-1}\times\sqrt{\overline{[\xi_{W}^2]^2}-\left(\overline{[\xi_{W}^2]}\right)^2},
\end{equation}
where the second moment of the jackknife replicates is given by $\overline{[\xi_{W}^2]^2}=\frac{1}{M/2}\sum_{i=1}^{M/2}\left[\xi_{W}^2\left([x]_i,[x^2]_i,[y]_i\right)\right]^2$.
This error bar scales as \( O(1/\sqrt{M/2}) \), substantially exceeding the bias in the estimate (see Supplemental Material Sec.~I).

\begin{figure*}[ht]
    \centering
    \begin{minipage}{\textwidth} 
     \centering
        \begin{tikzpicture}
            \node[anchor=south west,inner sep=0] (image) at (0,0) {\includegraphics[width=0.8\textwidth]{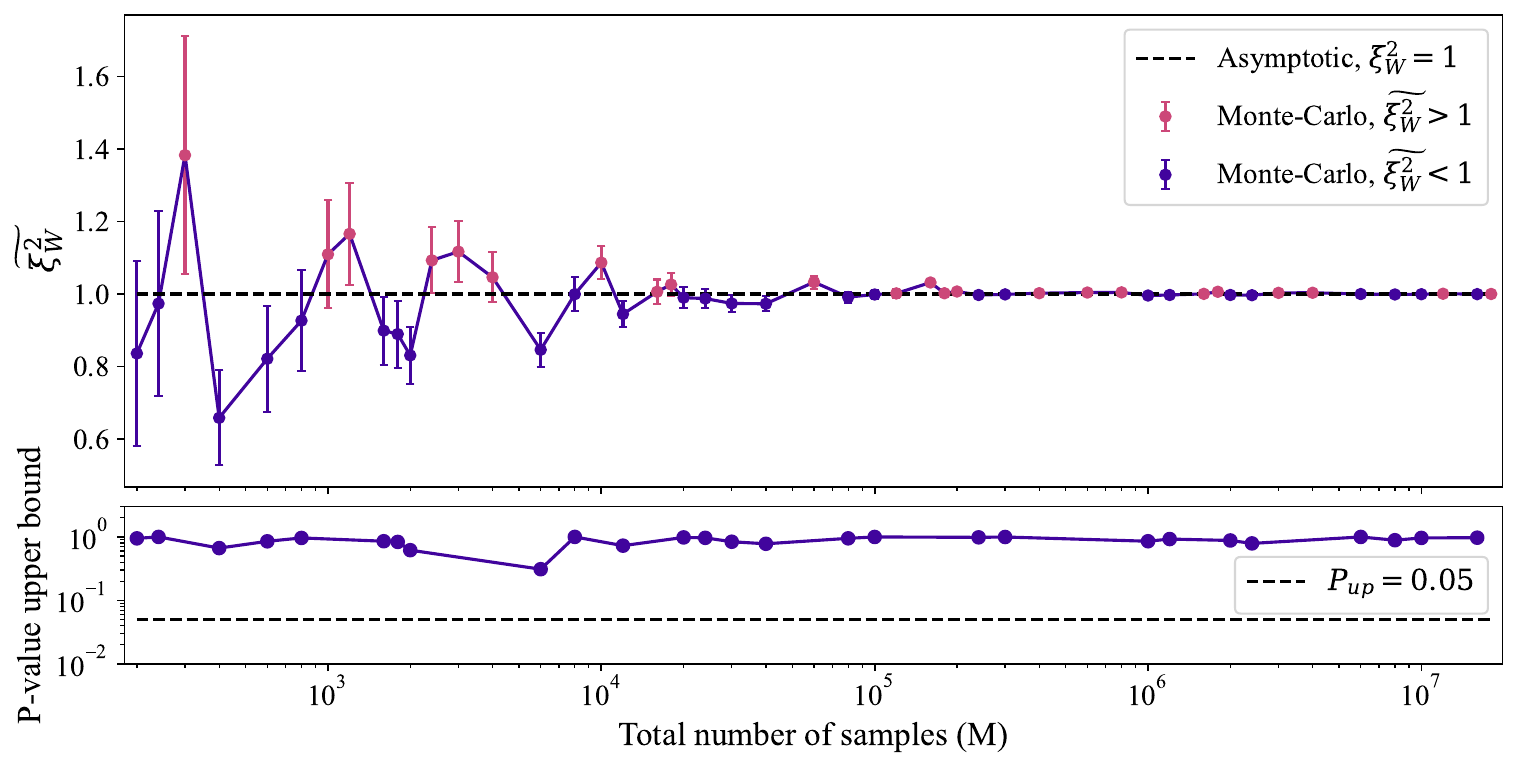}};
        \end{tikzpicture}
    \end{minipage}
    \caption{(Top) Estimation of the Wineland parameter from Monte Carlo samples of size \(M\) sampled from a non-spin-squeezed state. Two data sets were sampled: one in the eigenbasis of the mean spin direction (\(J_y\)) and another in the eigenbasis of the squeezing axis. Each error bar corresponds to one standard deviation.  (Bottom) P-value upper bound using the method proposed in the main text on the Monte-Carlo samples.}
    \label{misclassification}
\end{figure*}

\smallskip
\noindent
Figure~\ref{misclassification} shows the results of the Monte Carlo simulation with the jackknife estimation and uncertainty with one standard error, i.e.
\begin{align}
\label{s31}
\widetilde{\xi^2_W} &=   (M/2) \xi_{W}^2\left(\bar{x}, \overline{x^2}, \bar{y}\right) - (M/2-1) \overline{[\xi_{W}^2]} \\ \nonumber
&\pm  \sqrt{M/2-1}\times\sqrt{\overline{[\xi_{W}^2]^2}-\left(\overline{\xi_{W}^2}\right)^2}
\end{align}
as a function of the total number of samples $M$. Although these values fluctuate around the true value $\xi^2_W=1$, some are noticeably below 1. For instance, for $M=400$, we observe $\widetilde{\xi^2_W}=0.65832\pm0.13115$, despite the data being generated by a non-spin-squeezed state. The same figure also displays the p-value as a function of the number of measurements. None of the p-value drops below 0.25037. In particular, the p-value for $M=400$ is below 0.66661 clearly indicating a high probability that the measured state is non-spin-squeezed. More generally, none of these results can pass the hypothesis test and thus do not provide statistically significant evidence for spin squeezing. These simulation results (see also the complementary results presented in Supplemental Material Sec.~I) emphasize that ``error bars primarily quantify the spread of the data or the precision of the mean estimate, but do not constitute a rigorous test for statistical significance~\cite{Krzywinski2013}". Therefore, to provide a more reliable assessment of spin squeezing, it is essential to incorporate p-values as a complementary statistical measure to properly evaluate the probability that a non-spin squeezed state reproduces the observed statistics.}}

\end{document}

\title{Supplemental Material: Testing the Wineland Criterion with Finite Statistics}
%\title{Have we gone below the standard quantum limit on spin {\color{red}{squeezed}} systems?}% Force line breaks with \\
\author{E. S. Carrera}
\thanks{These authors contributed equally to this work}
%\email{ecarrera@ipht.fr}

\author{Y. Zhang}
\thanks{These authors contributed equally to this work}
%\email{yuzhang@ipht.fr}
\author{J-D. Bancal}
%\email{jbancal@ipht.fr}
\author{N. Sangouard}
%\email{nicolas.sangouard@ipht.fr}
\affiliation{%
 Université Paris-Saclay, CEA, CNRS, Institut de physique théorique, 91191, Gif-sur-Yvette, France}%

\date{\today}% It is always \today, today,

\maketitle

\section{Misclassification of Non-Spin-Squeezed States Using Error Bar Methods}

{We here review standard error bar estimation methods and discuss in particular, how they apply to the Wineland parameter. Importantly, we emphasize that the error bar solely reflects the statistical spread of the data and does not quantify the likelihood that the observed data could be reproduced by non-spin-squeezed states. This is illustrated in {End Matter} from the statistics of a non-spin-squeezed state obtained with Monte Carlo techniques by explicitly showing that small error bars do not necessarily correspond to small p-values. This underscores the risk of misclassifying a non-spin-squeezed state as spin-squeezed when relying on error bars to assess the significance of experimental results.} 

\subsection{Statistics of a non spin-squeezed state}
{We consider the statistics that would be produced from states of the form 
\begin{equation}
    \hat{\rho} = r \ket{\chi}\bra{\chi} + \dfrac{1-r}{2}\left( \ket{\pmb{\uparrow_{\Vec{n}_{{\perp}}}}}\bra{\pmb{\uparrow_{\Vec{n}_{{\perp}}}}} + \ket{\pmb{\downarrow_{\Vec{n}_{{\perp}}}}}\bra{\pmb{\downarrow_{\Vec{n}_{{\perp}}}}}\right)
\end{equation}
where $\ket{\chi}$ is a squeezed state with mean spin direction $\Vec{n}$, $\ket{\pmb{\uparrow_{\Vec{n}_{{\perp}}}}}=\ket{\uparrow_{\Vec{n}_{{\perp}}}}^{\otimes N} $, $\ket{\pmb{\downarrow_{\Vec{n}_{{\perp}}}}}=\ket{\downarrow_{\Vec{n}_{{\perp}}}}^{\otimes N}$ and $0\leq r\leq 1$. Inspired by two-axis twisting methods recently applied in Rydberg atom quantum simulators, $|\chi\rangle$ is generated from the time evolution of a coherent spin state under a dipolar XY model Hamiltonian
\begin{equation}
    H_{\text{XY}} = J\sum_{i < j} \frac{a^3}{r_{ij}^3} \left( S_i^x S_j^x + S_i^y S_j^y \right),
\end{equation}
where $S_i^\alpha$ are the spin-1/2 Pauli operators, $r_{ij}$ is the distance between spin $i$ and $j$, $J$ denotes the dipolar interaction strength, and $a$ is the lattice spacing. We consider in particular, a square array of $N=9$ spin-1/2 particles, initially prepared in a coherent spin-squeezed state aligned along the $y$-axis. The result of a time-evolution simulation using the library \textit{DynamiQs}~\cite{guilmin2024dynamiqs} shows a spin-squeezed state with squeezing direction $\vec{x}$ and mean spin direction $\vec{y}$ characterized by $\xi_{W}^{2}(\ket{\chi}) \approx 0.41036$, $\langle 2J_y/N\rangle_{\chi} \approx 0.80812$ and $\text{Var}\left(2 J_{x}/N\right)_{\chi} \approx 0.02977$ at time $t \approx 0.20707/J$. Using this state, we construct a non-spin-squeezed state satisfying $\xi^2_{W}(\rho) = 1$, which is characterized by $r\approx 0.96154$, $\langle 2J_y/N\rangle_{\rho} \approx 0.77704$ and  $\text{Var}\left(2 J_{x}/N\right)_{\rho} \approx 0.06709$.}

\medskip
\noindent
{The statistics that would be obtained from such a state are generated using Monte Carlo sampling.
Two data sets, each of size $M/2$, are produced ($M$ is the total number of samples). The first data set, $Q_{y}=\{y_1, \dots, y_{M/2}\}$, is sampled in the eigenbasis of $2J_{y}/N$. The second data set, $Q_{x}=\{x_1, \dots, x_{M/2}\}$, is sampled in the eigenbasis of $2J_{x}/N$. To establish
a comprehensive error estimation framework, the measurement outcomes $x_i$ and $y_j$ are assumed to be independent and
identically distributed (i.i.d.).}
%
%{\textcolor{red}{Are the names "population expectation/variances" usual? Where did you find them?}}\textcolor{olive}{We have replaced the "population expectation/variances" by "true expectation/variances"}
%\textcolor{red}{I still see the words "population expectation/variances". Can you please remove them all?} \textcolor{olive}{Checked!}

\medskip
\noindent
{Although the underlying probability distributions of  $x_i$ and $y_i$ are generally unknown, we define their true expectations as $\mu_x=\langle x_i\rangle$ and $\mu_y=\langle y_i\rangle$, and their true variances as
$\sigma^2_x=\langle (x_i-\mu_x)^2\rangle$ and $\sigma^2_y=\langle (y_j-\mu_{y})^2\rangle$. Here, the notation $ \langle\cdot\cdot\cdot\rangle$
represents the true expectation over the unknown underlying distribution, which is a quantity that cannot be directly observed in practice. In our analysis, we further consider the second-order moment of $x_i$ and used $\mu_{x^2}=\langle x_{i}^{2}\rangle$ to label its true expectation. Note that $\mu_{x^2}=\sigma_{x}^2+\mu_{x}^2$. With these quantities at hand, the Wineland parameter $ \xi_W^2 $ can be expressed as
\begin{equation}
\label{s3}
\xi_W^2 = \frac{N[\mu_{x^2}-(\mu_{x})^2]}{(\mu_{y})^2}.
\end{equation}
}

\medskip
\noindent
{To estimate the Wineland parameter from a given dataset $Q_{x}$ and $Q_{y}$, it is essential to obtain empirical estimates of $\mu_{x}$, $\mu_{x^2}$ and $\mu_{y}$. Since the true means and variances of the underlying distributions are not directly accessible, we estimate them using their sample counterparts, which are computed from a finite number of observations. We begin by defining the sample means of the data sets $Q_{x}$ and $Q_{y}$, denoted as $\bar{x}$ and $\bar{y}$, respectively. Here, the overbar notation $\overline{\cdot\cdot\cdot}$ represents the sample mean computed from $M/2$ data points. In addition, we  introduce the sample variance $s_{x}^{2}$ and $s_{y}^{2}$, which quantify the spread of the data, as well as the second raw moment of $Q_{x}$ (mean square), denoted as $ \overline{x^2}$. These quantities are given by
\begin{align}
    &\overline{x}=\frac{1}{M/2}\sum_{i=1}^{M/2}x_{i},\quad s^{2}_{x}=\frac{1}{M/2-1}\sum_{i=1}^{M/2}(x_i-\bar{x})^{{2}}, \quad \overline{x^{2}}=\frac{1}{M/2}\sum_{i=1}^{M/2}x_{i}^{2}
    \\ \nonumber 
    &\overline{y}=\frac{1}{M/2}\sum_{i=1}^{M/2}y_{i},\quad s^{2}_{y}=\frac{1}{M/2-1}\sum_{i=1}^{M/2}(y_i-\bar{y})^{{2}}.
\end{align}
We have
\begin{align}
&\langle\overline{x}\rangle=\frac{1}{M/2}\sum_{i=1}^{M/2}\langle x_i\rangle=\mu_x, \quad \sigma^{2}_{x}=\langle s^2_{x} \rangle, \quad \langle \overline{x^2}\rangle=\frac{1}{M/2}\sum_{i=1}^{M/2}\langle x_i^2 \rangle=\sigma_{x}^{2}+\mu_{x}^{2} =\mu_{x^2},\\ \nonumber
&\langle\overline{y}\rangle=\frac{1}{M/2}\sum_{i=1}^{M/2}\langle y_i\rangle=\mu_y, \quad \sigma^{2}_{y}=\langle s^2_{y} \rangle.
\end{align}
These results indicate that the sample means $\bar{x}$, $\overline{x^2}$ and $\bar{y}$ serve as unbiased estimators of their respective true expectations $\mu_x$, $\mu_{x^2}$ and $\mu_{y}$. Likewise, the sample variances $s_{x}^2$ and $s_{y}^2$ provide unbiased estimators of the true variances $\sigma_{x}^2$ and $\sigma_{y}^2$, ensuring that, on average, these estimates converge to their theoretical counterparts as the sample size increases.}

\medskip
\noindent
{Using these empirical quantities, the most natural approach to estimate $\xi_{W}^2$ is to approximate the true expectation by their sample counterparts, namely, setting $\mu_x\approx\bar{x},\mu_y\approx\bar{y}$ and $\mu_{x^2}\approx\overline{x^2}$. %This procedure is theoretically well-motivated and commonly adopted in experimental analyses. 
However, due to the nonlinear nature of $\xi_{W}^2$, this direct substitution will introduce a systematic bias in the estimation~\cite{becca2017quantum}. To mitigate this issue, experimental studies often employ statistical techniques such as error propagation~\cite{Bohnet2016}, the bootstrap method, or the jackknife method~\cite{Franke2023,Strobel2014}
%\textcolor{red}{Do you have similar citations for error propagation and bootstrap, i.e. experiments reporting on spin squeezing using these methods?} \textcolor{olive}{Only three papers mention the method for calculating the error bar, but none provide a detailed procedure.}
to assess and correct for biases introduced by nonlinear transformations.  In this work, we focus on the jackknife resampling technique, which will be introduced in subsequent sections. This method provides an effective means of quantifying the bias arising from finite sample effects and improving the accuracy of the estimated Wineland parameter.}

\subsection{Estimating the Wineland parameter with the Jackknife Method}

{The jackknife resampling method~\cite{becca2017quantum} is a widely used statistical technique for estimating uncertainties in nonlinear functions. It provides a systematic and fully automated approach to determining both error bars and bias corrections.}

\begin{proposition}
\label{prop1}{Consider three independent datasets, each consisting of $L$ i.i.d. random variables, $ \{u_1, \dots, u_L\}$, $ \{v_1, \dots, v_L\}$ , and $\{w_1, \dots, w_L\} $. The corresponding true means and variances are denoted as $ \mu_u, \mu_v, \mu_w $ and $ \sigma_u^2, \sigma_v^2, \sigma_w^2 $, respectively. Let the sample means of the datasets be denoted as $\bar{u}=\frac{1}{L}\sum_{i=1}^{L}{u_i}$, $\bar{v}=\frac{1}{L}\sum_{i=1}^{L}{v_i}$ and $\bar{w}=\frac{1}{L}\sum_{i=1}^{L}{w_i}$. Let $\{[u]_i\}, \{[v]_i\}$, and $\{[w]_i\}$ be the Jackknife datasets, where $[\cdot]_{i}$ is the jackknife averages computed by systematically leaving out the $i$-th data point from the original dateset, i.e., $[u]_{i}=\frac{1}{L}\sum_{j\neq i}u_{j}, [v]_{i}=\frac{1}{L}\sum_{j\neq i}v_{j}, [w]_{i}=\frac{1}{L}\sum_{j\neq i}w_{j}$.  Let $\overline{[\cdot]}=\frac{1}{L}\sum_{i=1}^{L}[\cdot]_{i}$ be the sample mean of the jackknife averages, and $\overline{[\cdot]^2} = \frac{1}{L} \sum_{i=1}^{L} [\cdot]_i^{2}$ the second raw moment.
For any differentiable function $f$, denoting the $i$-th jackknife estimates of $f$ as  $ [f]_i = f([u]_i, [v]_i, [w]_i)$, the jackknife estimate of  $f(\mu_u, \mu_v, \mu_w)$ is then given by
\begin{equation}
    L f(\bar{u}, \bar{v}, \bar{w}) - (L-1) \overline{[f]},
\end{equation}
%\textcolor{red}{$\bar u$ is not defined. Similarly for $\bar v$ and $\bar w$. By the way, do you mean $\overline{[u]}$?}\textcolor{olive}{For a single-variable or linear quantity, the sample mean remains unchanged under the jackknife resampling process, satisfying the relation $\bar{\cdot}=\overline{[\cdot]}$. However, for nonlinear quantities, this equivalence no longer holds. }
where $\overline{[f]}=\frac{1}{L}\sum_{i=1}^{L}[f]_{i}$ is the mean of jackknife estimates.  For the variance of $f(\mu_u, \mu_v, \mu_w)$, it is estimated by
\begin{equation}
    (L-1)\left(\overline{[f]^2}-\left( \overline{[f]} \right)^2\right),
\end{equation}
where \( \overline{[f]^2}-\left( \overline{[f]} \right)^2=s_{[f]}^2\) is the jackknife variance of the jackknife estimates. %\textcolor{red}{Please rewrite the previous sentence. It is currently unclear. Is S-19 the definition of \( s_{[f]}^{2} \)?}  
The bias of the jackknife estimate of $f(\mu_u, \mu_v, \mu_w)$ scales as $O(1/L)$, while the estimated error bar $\sqrt{(L-1) s_{[f]}^{2}}$ scales as $O(1/\sqrt{L})$.}
\end{proposition}

\begin{proof}
{We begin by identifying the leading-order contributions to the bias in the estimation of $f(\mu_u, \mu_v, \mu_w)$. Due to the nonlinear nature of $f$, direct substitution of the true means by the sample means generally introduces a systematic bias. To quantify this effect, we expand $f(\bar{u},\bar{v},\bar{w})$ around its expectation at the true mean values, $ f(\mu_u, \mu_v, \mu_w)$, using a multivariate Taylor series expansion, retaining terms up to the second order. We get
\begin{align}
\label{s8}
f(\bar{u},\bar{v},\bar{w})= &f(\mu_{u},\mu_{v},\mu_{w}) + (\partial_{\mu_{u}}{f})\Delta_{u} + (\partial_{\mu_{v}}{f})\Delta_{v} + (\partial_{\mu_{w}}f{})\Delta_{w} \\ \nonumber
&+\frac{1}{2}(\partial_{\mu_{u},\mu_{u}}^{2}{f})\Delta_{u}^{2}+ \frac{1}{2}(\partial_{\mu_{v},\mu_{v}}^{2}{f})\Delta_{v}^{2}+\frac{1}{2}(\partial_{\mu_{w},\mu_{w}}^{2}{f})\Delta_{w}^{2} \\ \nonumber
&+(\partial_{\mu_{u},\mu_{v}}^{2}{f})\Delta_{u}\Delta_{v}+(\partial_{\mu_{u},\mu_{w}}^{2}{f})\Delta_{u}\Delta_{w} +(\partial_{\mu_{v},\mu_{w}}^{2}{f})\Delta_{v}\Delta_{w},
\end{align}
where \( \Delta_{u}=\bar{u}-\mu_{u} \), \( \Delta_{v}=\bar{v}-\mu_{v} \) and \( \Delta_{w}=\bar{w}-\mu_{w} \). The leading contribution to the bias comes
from the second-order terms, since the first-order terms in \( \Delta_{u} \), \( \Delta_{v} \), and \( \Delta_{w} \) have
zero expectation, i.e., \( \langle\Delta_{u}\rangle=0 \), \( \langle\Delta_{v}\rangle=0 \), and \( \langle\Delta_{w}\rangle=0 \). Instead,
the second-order terms have finite expectation values
\begin{align} &\langle\Delta_{u}^{2}\rangle=\langle\bar{u}^{2}\rangle - \langle\bar{u}\rangle^2 = \sigma_{\bar{u}}^2=\frac{\sigma_{u}^2}{L}, \quad \langle\Delta_{v}^{2}\rangle=\langle\bar{v}^{2}\rangle - \langle\bar{v}\rangle^2 = \sigma_{\bar{v}}^2=\frac{\sigma_{v}^2}{L}, \quad \langle\Delta_{w}^{2}\rangle=\langle\bar{w}^{2}\rangle - \langle\bar{w}\rangle^2 = \sigma_{\bar{w}}^2=\frac{\sigma_{w}^2}{L}, \\ \nonumber
&\langle\Delta_{u}\Delta_{v}\rangle=\langle\bar{u}\bar{v}\rangle - \langle\overline{u}\rangle \langle\overline{v}\rangle = \sigma_{\bar{u}\bar{v}}^2=\frac{\sigma_{uv}^2}{L}, \quad \langle\Delta_{u}\Delta_{w}\rangle=\langle\bar{u}\bar{w}\rangle - \langle\bar{u}\rangle \langle\bar{w}\rangle = \sigma_{\bar{u}\bar{w}}^2=\frac{\sigma_{uw}^2}{L}, \\ \nonumber
&\langle\Delta_{v}\Delta_{w}\rangle=\langle\bar{v}\bar{w}\rangle - \langle\bar{v}\rangle \langle\bar{w}\rangle = \sigma_{\bar{v}\bar{w}}^2=\frac{\sigma_{vw}^2}{L}.
\end{align}
Here, $\sigma_{uv}$, $\sigma_{uw}$ and $\sigma_{vw}$ are the covariance. Substituting these expressions into the Taylor expansion, we obtain 
\begin{align}
\label{S-mean}
    \langle f(\bar{u},\bar{v},\bar{w})\rangle-f(\mu_{u},\mu_{v},\mu_{w})=&\frac{1}{2L}\left[(\partial_{\mu_{u},\mu_{u}}^{2}{f})\sigma_{u}^2+ (\partial_{\mu_{v},\mu_{v}}^{2}{f})\sigma_{v}^2+(\partial_{\mu_{w},\mu_{w}}^{2}{f})\sigma_{w}^2\right] \\ \nonumber
&+\frac{1}{L}\left[(\partial_{\mu_{u},\mu_{v}}^{2}{f})\sigma_{uv}^2+(\partial_{\mu_{u},\mu_{w}}^{2}{f})\sigma_{uw}^2 +(\partial_{\mu_{v},\mu_{w}}^{2}{f})\sigma_{vw}^2\right],
\end{align}
which proves that the leading bias in estimating \( f(\mu_{u},\mu_{v},\mu_{w}) \) from \( f(\bar{u},\bar{v},\bar{w}) \) is of order \( O(1/L) \). }
%
%\medskip
%\noindent
\begin{comment}
\sout{Next, we analyze the leading contribution to the statistical uncertainty in \(f(\bar{u},\bar{v},\bar{w}) \). \textcolor{blue}{With the expression~\ref{s8}}, the variance of the estimate can be computed as:}
% \begin{align}
% \label{Svar}
%     \sigma_{f}^{2}&=\langle f(\bar{u},\bar{v},\bar{w})^2\rangle-\langle f(\bar{u},\bar{v},\bar{w})\rangle^{2} =\langle [ f(\bar{u},\bar{v},\bar{w})-\langle f(\bar{u},\bar{v},\bar{w})\rangle ]^2\rangle = \langle [f(\bar{u},\bar{v},\bar{w}) -f(\mu_u,\mu_v,\mu_w)]^2\rangle
%     \\ \nonumber
%     &=\frac{1}{M}\left[(\partial_{\mu_{u}}{f})^{2}\sigma_{u}^2+ (\partial_{\mu_{v}}{f})^{2}\sigma_{v}^2+(\partial_{\mu_{w}}{f})^{2}\sigma_{w}^2\right] +\frac{2}{M}\left[(\partial_{\mu_{u}}{f})(\partial_{\mu_{v}}{f})\sigma_{uv}^2+(\partial_{\mu_{u}}{f})(\partial_{\mu_{w}}{f})\sigma_{uw}^2 +(\partial_{\mu_{v}}{f})(\partial_{\mu_{w}}{f})\sigma_{vw}^2\right],
% \end{align}
\textcolor{red}{I don't immediately see how to get the very last term in the second line of S-11. Can you comment?}\textcolor{olive}{Using Eq. (S8), we first compute $[f(\bar{u},\bar{v},\bar{w})-f(\mu_{u},\mu_{v},\mu_{w})]^2$, and then take the expectation to obtain the desired result. During the calculation, we retain only the leading-order contributions, as higher-order terms contribute at the $O(1/M^2)$ level and can be neglected.}
\end{comment}
%

\medskip
\noindent
{We now establish the jackknife estimate for $f(\mu_u, \mu_v, \mu_w)$.}
\begin{comment}
\begin{equation}
    A_{\sigma}=\frac{1}{2}\left[(\partial_{\mu_{u},\mu_{u}}^{2}{f})\sigma_{u}^2+ (\partial_{\mu_{v},\mu_{v}}^{2}{f})\sigma_{v}^2+(\partial_{\mu_{w},\mu_{w}}^{2}{f})\sigma_{w}^2\right] +(\partial_{\mu_{u},\mu_{v}}^{2}{f})\sigma_{uv}^2+(\partial_{\mu_{u},\mu_{w}}^{2}{f})\sigma_{uw}^2 +(\partial_{\mu_{v},\mu_{w}}^{2}{f})\sigma_{vw}^2.
\end{equation}
\end{comment}
{We first extend \ref{S-mean} formally to get second order contributions, that is, we write
\begin{equation}
\label{s25}
    f(\mu_u,\mu_v,\mu_w) = \langle f(\bar{u},\bar{v},\bar{w}) \rangle - \frac{A_{\sigma}}{L} + \frac{B_{\sigma}}{L^2},
\end{equation}
where $A_{\sigma}/L$ and $B_{\sigma}/L^2$ are the first and second order contributions to the bias. From \ref{S-mean}, we have
\begin{equation}
    A_{\sigma}=\frac{1}{2}\left[(\partial_{\mu_{u},\mu_{u}}^{2}{f})\sigma_{u}^2+ (\partial_{\mu_{v},\mu_{v}}^{2}{f})\sigma_{v}^2+(\partial_{\mu_{w},\mu_{w}}^{2}{f})\sigma_{w}^2\right] +(\partial_{\mu_{u},\mu_{v}}^{2}{f})\sigma_{uv}^2+(\partial_{\mu_{u},\mu_{w}}^{2}{f})\sigma_{uw}^2 +(\partial_{\mu_{v},\mu_{w}}^{2}{f})\sigma_{vw}^2.
\end{equation}
}{To derive the jackknife estimator, we expand each leave-one-out estimate $[f]_{i}$ around \( f(\mu_u,\mu_v,\mu_w) \), 
\begin{align} 
\label{s26}
\overline{[f]}&=f(\mu_u,\mu_v,\mu_w)+(\partial_{\mu_u}f)(\bar{u}-\mu_u)+(\partial_{\mu_v}f)(\bar{v}-\mu_v)+(\partial_{\mu_w}f)(\bar{w}-\mu_w) \\ \nonumber
    &+\frac{(\partial_{\mu_u,\mu_u}^2f)}{2L}\sum_{i=1}^{L}([u]_{i}-\mu_u)^2 + \frac{(\partial_{\mu_v,\mu_v}^2f)}{2L}\sum_{i=1}^{L}([v]_{i}-\mu_v)^2 +\frac{(\partial_{\mu_w,\mu_w}^2f)}{2L}\sum_{i=1}^{L}([w]_{i}-\mu_w)^2 \\ \nonumber
    &+\frac{(\partial_{\mu_u,\mu_v}^2f)}{L}\sum_{i=1}^{L}([u]_{i}-\mu_u)([v]_{i}-\mu_v) + \frac{(\partial_{\mu_u,\mu_w}^2f)}{L}\sum_{i=1}^{L}([u]_{i}-\mu_u)([w]_{i}-\mu_w) + \frac{(\partial_{\mu_v,\mu_w}^2f)}{L}\sum_{i=1}^{L}([v]_{i}-\mu_v)([w]_{i}-\mu_w) \\ \nonumber
    &+ O\left(\frac{1}{L^2}\right).
\end{align}}
\begin{comment}
\textcolor{red}{In the second term of the first line of Eq. S-26, is it $\bar u$ or $\overline{[u]}$? Similar questions for the third and fourth term in the same line.}\textcolor{olive}{For the first line, it is $\bar{u}$. For the second and third line, it is not $\bar{u}$. This can be checked by performing simple calculation. For a single-variable or linear quantity, the sample mean remains unchanged under the jackknife resampling process, satisfying the relation, $\bar{\cdot}=\overline{[\cdot]}$. During the computation process, whenever possible, quantities should be expressed in terms of $\bar{u}$, as it represents the fundamental and directly observable sample mean.}
\end{comment}
{For the sums, we use the following identities
\begin{align}
    \frac{1}{L}\sum_{i=1}^{L}([u]_{i}-\mu_u)^2&=(\bar{u}-\mu_u)^2+\frac{s_{u}^2}{L(L-1)} \\ \nonumber
    \frac{1}{L}\sum_{i=1}^{L}([u]_{i}-\mu_u)([v]_{i}-\mu_v)&=(\bar{u}-\mu_u)(\bar{v}-\mu_{v})+\frac{s_{uv}^2}{L(L-1)}.
\end{align}
Here, $s_{uv}^2$ denotes the sample covariance
\begin{equation}
    s_{uv}^2=\frac{1}{L-1}\sum_{i=1}^{L}(u_i-\bar{u})(v_i-\bar{v}),
\end{equation}
which is an unbiased estimator for the covariance $\sigma_{uv}^2$.
Similar expressions hold for the other summation terms. Substituting these and taking the expectation of \ref{s26}, the jackknife expansion is given by
\begin{equation}
\label{s27}
     f(\mu_u,\mu_v,\mu_w) = \langle \overline{[f]} \rangle - \frac{A_{\sigma}}{L-1}+\frac{B_{\sigma}^{'}}{(L-1)^2}.
\end{equation}
Here, the next order correction is represented by $B_{\sigma}^{'}$, which is different from $B_{\sigma}$ due to the jackknife averages. We can therefore eliminate the leading contribution to the bias by forming an appropriate linear combination of \ref{s25} and \ref{s27}, that is
\begin{equation}
    f(\mu_u,\mu_v,\mu_w)= L \langle f(\bar{u},\bar{v},\bar{w}) \rangle - (L-1) \langle \overline{[f]} \rangle + \frac{B_{\sigma}}{L}-\frac{B_{\sigma}^{'}}{L-1}.
\end{equation}
%Noted that the next order correction ${B_{\sigma}}/{M}-{B_{\sigma}^{'}}/{(M-1)}$ is of order \( O(1/M) \), it is significantly smaller than the statistical error which scales as $O(1/\sqrt{M})$ (as we will demonstrate later). 
Consequently, to eliminate the leading bias without computing partial derivatives, we estimate $f(\mu_u,\mu_v,\mu_w)$ by
\begin{equation}
    L  f(\bar{u},\bar{v},\bar{w}) - (L-1)  \overline{[f]}.
\end{equation}
}

\medskip
\noindent
{Next, we analyze the leading contributions to the statistical uncertainty. The variance $\sigma_{f}^2$ of the estimate  \(f(\bar{u},\bar{v},\bar{w}) \) is  defined as~\cite{becca2017quantum}
\begin{align}
\label{Svar}
    \sigma_{f}^{2}&=\langle f(\bar{u},\bar{v},\bar{w})^2\rangle-\langle f(\bar{u},\bar{v},\bar{w})\rangle^{2} =\langle [ f(\bar{u},\bar{v},\bar{w})-\langle f(\bar{u},\bar{v},\bar{w})\rangle ]^2\rangle = \langle [f(\bar{u},\bar{v},\bar{w}) -f(\mu_u,\mu_v,\mu_w)]^2\rangle
    \\ \nonumber
    &=\frac{1}{L}\left[(\partial_{\mu_{u}}{f})^{2}\sigma_{u}^2+ (\partial_{\mu_{v}}{f})^{2}\sigma_{v}^2+(\partial_{\mu_{w}}{f})^{2}\sigma_{w}^2\right] +\frac{2}{L}\left[(\partial_{\mu_{u}}{f})(\partial_{\mu_{v}}{f})\sigma_{uv}^2+(\partial_{\mu_{u}}{f})(\partial_{\mu_{w}}{f})\sigma_{uw}^2 +(\partial_{\mu_{v}}{f})(\partial_{\mu_{w}}{f})\sigma_{vw}^2\right].
\end{align}
where, to get the last line, we omitted all second-order terms in~\ref{s8} as they contribute at the $O(1/L^2)$ level. %To quantify the variance $\sigma_{f}^2$, it requires the exact calculation of all partial derivatives, besides keeping track of all the variances and covariances.  
Interestingly, the jackknife method can estimate $\sigma_{f}^2$  without requiring explicit computation of partial derivatives. %, and the resulting error bar is of order \( O(1/\sqrt{M}) \). 
To demonstrate this, we expand each \( [f]_{i} \) around \( f(\mu_u,\mu_v,\mu_w) \) and include only the leading contribution, which gives,  
\begin{equation}
    \overline{[f]}-f(\mu_u,\mu_v,\mu_w)=(\partial_{\mu_u}f)(\bar{u}-\mu_u)+(\partial_{\mu_v}f)(\bar{v}-\mu_v)+(\partial_{\mu_w}f)(\bar{w}-\mu_w).
\end{equation}
%\textcolor{red}{Do you need the non-leading terms of S-20 for obtaining S-22?}\textcolor{olive}{We don't need the non-leading terms.}
Similarly, for the second moment, we have
\begin{align}
\overline{[f]^2}&=\frac{1}{L}\sum_{i=1}^{L}\left[f(\mu_u,\mu_v,\mu_w)+(\partial_{\mu_u}f)([u]_{i}-\mu_u) +(\partial_{\mu_v}f)([v]_{i}-\mu_v)+(\partial_{\mu_w}f)([w]_{i}-\mu_w) \right]^2 \\ \nonumber
    &=f^2(\mu_u,\mu_v,\mu_w) + 2f(\mu_u,\mu_v,\mu_w)\left[(\partial_{\mu_u}f)(\bar{u}-\mu_u)+(\partial_{\mu_v}f)(\bar{v}-\mu_v)+(\partial_{\mu_w}f)(\bar{w}-\mu_w)\right] \\ \nonumber
    &+(\partial_{\mu_u}f)^2\left[(\bar{u}-\mu_u)^2+\frac{s_{u}^2}{L(L-1)}\right]+(\partial_{\mu_v}f)^2\left[(\bar{v}-\mu_v)^2+\frac{s_{v}^2}{L(L-1)}\right] +(\partial_{\mu_w}f)^2\left[(\bar{w}-\mu_w)^2+\frac{s_{w}^2}{L(L-1)}\right] \\ \nonumber
    &+2(\partial_{\mu_u}f)(\partial_{\mu_v}f)\left[(\bar{u}-\mu_u)(\bar{v}-\mu_v)+\frac{s_{uv}^2}{L(L-1)}\right] + 2(\partial_{\mu_u}f)(\partial_{\mu_w}f)\left[(\bar{u}-\mu_u)(\bar{w}-\mu_w)+\frac{s_{uw}^2}{L(L-1)}\right] \\ \nonumber
    &+2(\partial_{\mu_v}f)(\partial_{\mu_w}f)\left[(\bar{v}-\mu_v)(\bar{w}-\mu_w)+\frac{s_{vw}^2}{L(L-1)}\right].
\end{align}
Hence, the variance in the jackknife estimates is given by
\begin{align}
s_{[f]}^{2}=\overline{[f]^2}-\left( \overline{[f]} \right)^2 &=\frac{1}{L(L-1)}\left[(\partial_{\mu_{u}}{f})^{2}s_{u}^2+ (\partial_{\mu_{v}}{f})^{2}s_{v}^2+(\partial_{\mu_{w}}{f})^{2}s_{w}^2\right] \\ \nonumber
&+\frac{2}{L(L-1)}\left[(\partial_{\mu_{u}}{f})(\partial_{\mu_{v}}{f})s_{uv}^2+(\partial_{\mu_{u}}{f})(\partial_{\mu_{w}}{f})s_{uw}^2 +(\partial_{\mu_{v}}{f})(\partial_{\mu_{w}}{f})s_{vw}^2\right].
\end{align}
Taking the expectation and comparing with the variance given in Eq. \ref{Svar}, we obtain
\begin{equation}
    \langle s_{[f]}^2 \rangle = \frac{\sigma_{f}^2}{L-1}.
\end{equation}
Therefore, the jackknife estimate for \( \sigma_{f}^2 \) is given by \( (L-1)s_{[f]}^{2} \). Interestingly, the error bar $\sqrt{(L-1) s_{[f]}^{2}}$ scales as $O(1/\sqrt{L})$, while the
the first order correction of the correected bias is given by ${B_{\sigma}}/{L}-{B_{\sigma}^{'}}/{(L-1)}$ and is thus of the order of $O(1/L)$.
}
%\textcolor{red}{$s_{f^{J}}^{2}$ is not defined.}\textcolor{olive}{It has been replaced by $s_{[f]}^{2}$.}
\end{proof}
%\textcolor{red}{Are you ok with this version of the proof? Is something missing?}\textcolor{olive}{It is OK for me!}

\medskip
\noindent
{With the framework established, we now proceed with the estimation of \( \xi_W^2 = \xi_W^2(\mu_x, \mu_{x^2}, \mu_y) \). Since the data size of each date set, $Q_x$ and $Q_y$, is $M/2$,  Proposition~\ref{prop1} gives us the jackknife estimate of \( \xi_W^2 \)
\begin{equation}
    (M/2) \xi_{W}^2\left(\bar{x}, \overline{x^2}, \bar{y}\right) - (M/2-1) \overline{[\xi_{W}^2]}, \quad \text{where} \quad \overline{[\xi_{W}^2]}=\frac{1}{M/2}\sum_{i=1}^{M/2}\xi_{W}^2\left([x]_i,[x^2]_i,[y]_i\right)
\end{equation}
and tells us that the bias in this estimate is of order \( O(2/M) \). Here, $[x]_i,[x^2]_i,[y]_i$ are  jackknife averages 
\begin{align}
    [x]_{i}=\frac{1}{M/2-1}\sum_{j\neq i}x_i, \quad [x^2]_{i}=\frac{1}{M/2-1}\sum_{j\neq i}x_{i}^{2}, \quad [y]_{i}=\frac{1}{M/2-1}\sum_{j\neq i}y_i.
\end{align}
Proposition~\ref{prop1} also gives us
the uncertainty in the estimation of \( \xi_W^2(\mu_x, \mu_{x^2}, \mu_y) \). In particular, it provides the following estimate for the error bar 
\begin{equation}
    \sqrt{M/2-1}\times\sqrt{\overline{[\xi_{W}^2]^2}-\left(\overline{\xi_{W}^2}\right)^2}, \quad \text{where} \quad \overline{[\xi_{W}^2]^2}=\frac{1}{M/2}\sum_{i=1}^{M/2}\left[\xi_{W}^2\left([x]_i,[x^2]_i,[y]_i\right)\right]^2
\end{equation}
The estimated error bar scales as \( O(1/\sqrt{M/2}) \), which is significantly larger than the bias in the estimate of \( \xi_W^2 \). This ensures that the jackknife approach provides a reliable uncertainty quantification while effectively correcting for systematic biases in the estimation process.}

\subsection{The complementary simulation results}
\begin{figure*}[ht]
    \centering
    \begin{minipage}{\textwidth} 
     \centering
        \begin{tikzpicture}
            \node[anchor=south west,inner sep=0] (image) at (0,0) {\includegraphics[width=0.8\textwidth]{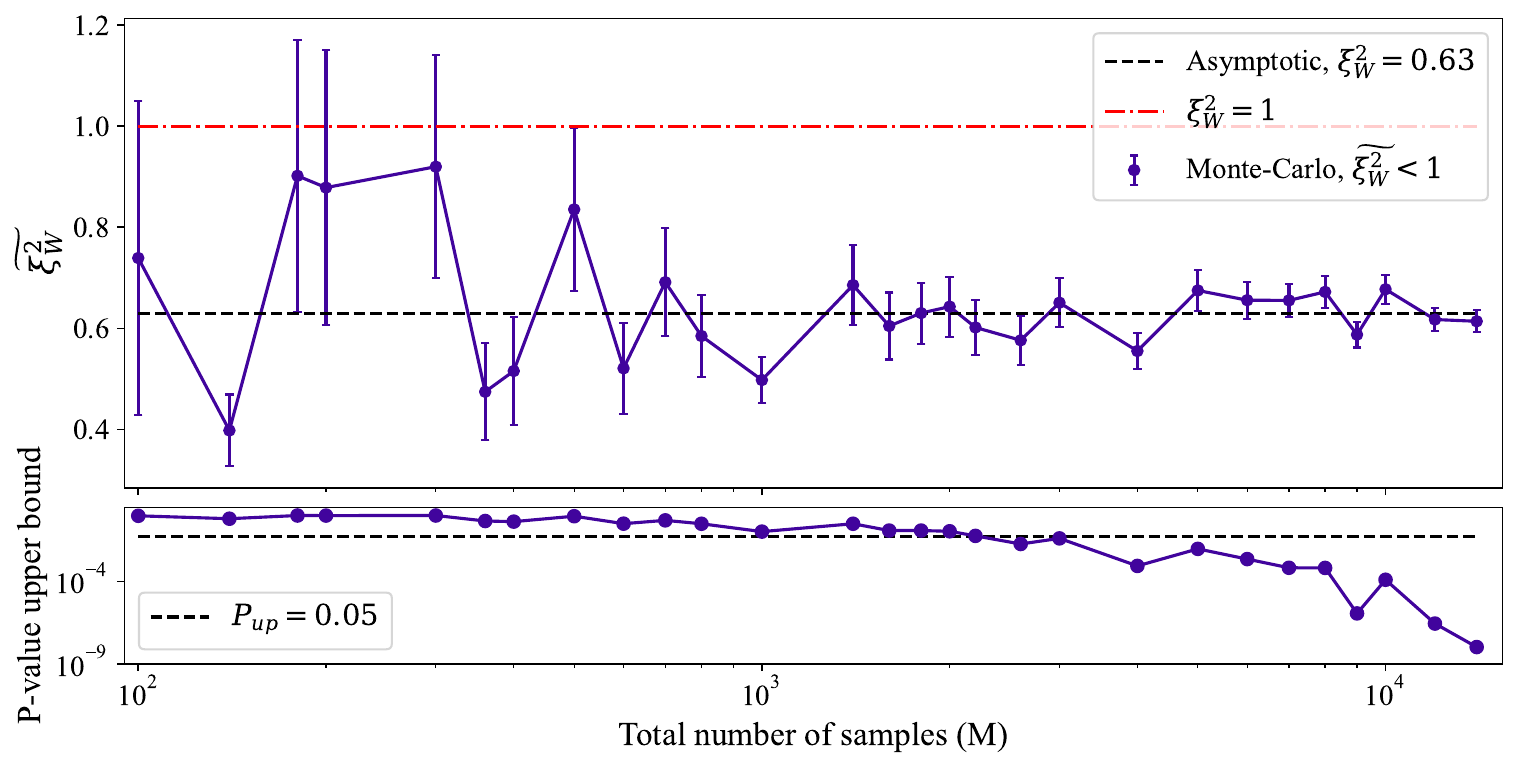}};
        \end{tikzpicture}
    \end{minipage}
    \caption{(Top) Estimation of the Wineland parameter from Monte Carlo samples of size \(M\) sampled from a spin-squeezed state. Two data sets were sampled: one in the eigenbasis of the mean spin direction (\(J_y\)) and another in the eigenbasis of the squeezing axis. Each error bar corresponds to one standard deviation.  (Bottom) P-value upper bound using the method proposed in the main text on the Monte-Carlo samples.}
    \label{fig:method_example}
\end{figure*}

{In the End Matter, we perform Monte Carlo simulations on a specific non-spin-squeezed state to demonstrate that small error bars do not necessarily imply low p-values. Conversely, for spin-squeezed states, the p-value can be used to assess the statistical significance of spin-squeezing generation experiments. To illustrate this, we generate Monte Carlo samples following the previously described state but with $r=0.985$. This state is characterized by $\xi^2_W \approx 0.62967$, $\langle 2 J_x/N\rangle \approx 0.79600$, and $\text{Var}(2 J_x / N) \approx 0.04433$. The corresponding values for $\widetilde{\xi^2_W}$ together with the p-value are presented in Fig.~\ref{fig:method_example} as a function of the number of measurements.}

\smallskip
\noindent
{Unlike the non-spin-squeezed scenario discussed in the End Matter, we observe that when the underlying state is spin-squeezed, the p-value upper bound decreases significantly once the number of measurements is high enough. As shown in the figure, without applying this method to assess the confidence of the experimental results, one might incorrectly conclude—due to small error bars—that the state is spin-squeezed with an estimated value of \(\widetilde{\xi^2_W} = 0.39815 \pm 0.07112\) from $M=140$ experimental repetitions (the second simulation result in Figure~\ref{fig:method_example}). However, the computed upper bound on the p-value is \(P_{\text{up}} = 0.62904\), indicating that the number of samples is insufficient to support this claim.}

\medskip
\noindent
{This result highlights the necessity of rigorous statistical validation in spin-squeezing experiments. Notably, in our simulation process, no specific probability distribution was assumed, ensuring that the observed misclassification is not a consequence of an improper assumption. The simulation results emphasize that ``error bars primarily quantify the spread of the data or the precision of the mean estimate, but do not constitute a rigorous test for statistical significance~\cite{Krzywinski2013}". Therefore, to provide a more reliable assessment of spin squeezing, it is essential to incorporate p-values as a complementary statistical measure to properly evaluate the probability that a non-spin squeezed state reproduces the observed statistics.}

\section{The overly conservative estimation form the parameter $\Gamma$}
%\section{The nonlinearity of the linearized Wineland criteria $\Gamma$}
\label{sec1}
As discussed in the main text, we consider the linearized version of the Wineland criterion which is expressed as a function of the variance $\text{Var}(Q_{\vec{n}_{\perp}})$ and the mean $\langle Q_{\vec{n}}\rangle$ as
\begin{equation}
\label{linwine}
    \Gamma = N \text{Var}(Q_{\vec{n}_{\perp}}) - \langle Q_{\vec{n}}\rangle^2.
\end{equation}
$Q_{\vec{n}_{\perp}}$ and $Q_{\vec{n}}$ are the normalized  angular momentum operators. To estimate the value of $\Gamma$ experimentally, it is necessary to identify unbiased estimators for $\text{Var}(Q_{\vec{n}_{\perp}})$ and $\langle Q_{\vec{n}}\rangle^2$. The sample variance and sample mean serve as natural choices for these estimators.  Specifically, we use for  $\text{Var}(Q_{\vec{n}_{\perp}})$ the following sample variance
\begin{equation}
\widetilde{\text{Var} (Q_{\Vec{n}_{\perp}}) }  = \dfrac{1}{M_{\Vec{n}_{\perp}} -1 }\sum_{k=1}^{M_{\Vec{n}_{\perp}}} \left(q_{{\Vec{n}_{\perp}}}^{(k)}  - \dfrac{1}{M_{\Vec{n}_{\perp}}}\sum_{j=1}^{M_{\Vec{n_\perp}}} q_{\Vec{n}_{\perp}}^{(j)}\right)^{2}.
\end{equation}
Here, $M_{\Vec{n}_{\perp}}$ is the number of measurements in the direction $\Vec{n}_{\perp}$. The term  $q^{(k)}_{\vec{n}_{\perp}}$ ($q^{(k)}_{\vec{n}})$  corresponds to the outcome of the measurement operator $2J_{\Vec{n}}^{(k)}/N$ $(2J_{\Vec{n}_{{\perp}}}^{(k)}/N)$ for experimental round $k$. For the estimator of $\langle Q_{\vec{n}}\rangle^2$, we  use the second moment, which is given by 
\begin{align}
    \widetilde{\langle Q_{\Vec{n}}\rangle^2} &=  \frac{1}{M_{\Vec{n}}} \sum_{i=1}^{M_{\Vec{n}}} q_{\Vec{n}}^{(i)2} - \widetilde{\text{Var} (Q_{\Vec{n}}) } \\
    & =\frac{1}{M_{\Vec{n}}} \sum_{k=1}^{M_{\Vec{n}}} q_{\Vec{n}}^{(k)2} -\dfrac{1}{M_{\Vec{n}} -1 }\sum_{k=1}^{M_{\Vec{n}}} \left(q_{{\Vec{n}}}^{(k)}  - \dfrac{1}{M_{\Vec{n}}}\sum_{j=1}^{M_{\Vec{n}}} q_{\Vec{n}}^{(j)}\right)^{2}.
\end{align}
It is important to note that both $ \widetilde{\langle Q_{\Vec{n}}\rangle^2}$ and $\widetilde{\text{Var} (Q_{\Vec{n}_{\perp}}) }$ are random variables. Using these estimators, the unbiased estimate of the linearized Wineland parameter can be computed as
\begin{equation} \widetilde{\Gamma}=N\widetilde{\text{Var} (Q_{\Vec{n}_{\perp}}) } -\widetilde{\langle Q_{\Vec{n}}\rangle^2}=g\left(q_{{\Vec{n}}}^{(1)},...,q_{{\Vec{n}}}^{(M_{\Vec{n}})},q_{{\Vec{n}_{\perp}}}^{(1)},...,q_{{\Vec{n}_{\perp}}}^{(M_{\Vec{n}_{\perp}})}  \right).
\end{equation}
It is important to note that the function $g$ used in the calculation of $\widetilde{\Gamma}$ cannot be expressed as an estimator of a sum of independent random variables. Instead, $g$ is a nonlinear function of these independent random variables. In this context, McDiarmid's inequality~\cite{McDiarmid1989,McDiarmid1998} and its various extensions are suited to determine the significance level. 

\medskip
\noindent
\textbf{McDiarmid's inequality:} 
McDiarmid's inequality applies to functions that satisfy the bounded differences property, which ensures that small changes in any single input result in only a limited change in the function’s output. Specifically, for a function $f$ of independent random variables $X_1, X_2, \dots, X_n$, the bounded differences property holds if for each i, $f$ changes by at most $s_i$ when substituting the value of the $i$th coordinate $x_i$ to $x_i'$, that is
\begin{equation}
  \forall i, \sup_{x_i' \in \mathcal{X}_i} \left|f(x_1, \dots, x_{i-1}, x_i, x_{i+1}, \ldots, x_n) - f(x_1, \dots, x_{i-1}, x_i', x_{i+1}, \ldots, x_n)\right| \leq s_i.
\end{equation}
McDiarmid's inequality then tells us that the deviation of the estimated value of $f$ from its expected value is bounded by
\begin{equation}
    \Pr\left(f(X_1, \dots, X_n) - \mathbb{E}[f(X_1, \dots, X_n)] \geq \epsilon \right) \leq \exp\left(\frac{-2\epsilon^2}{\sum_{i=1}^{n} s_i^2}\right).
\end{equation}
However, due to the bounded nature of $g$ and the significant magnitude of the bounded differences $s_i$ which scales as $O(N/M)$, applying McDiarmid's inequality results in a large upper bound on the p-value (see Section~\ref{other bounds} and  Fig.~\ref{allinone}). This large bound can lead to an overly conservative estimation when calculating statistical significance, effectively inflating the p-value and making it difficult to achieve strong conclusions regarding the experimental results. In Section~\ref{other bounds}, it is shown that by linearizing $\Gamma$, i.e., expressing it as a sum of independent random variables, the negative impact caused by the bounded differences can be effectively avoided.

\section{{The upper bound on the p-value from the parameter $\Gamma_{c}$}}\label{sec:bestBound}

To calculate the p-value associated to the estimator presented in Eqs. (5) and (6) of the main text, we use Bernstein’s inequality~\cite{McDiarmid1989,McDiarmid1998}, which is particularly suited for cases involving independent random variables with known variances. %Bernstein’s Inequality provides a probabilistic bound on the sum of independent random variables and is defined as follows,

\medskip
\noindent
{\textbf{Bernstein's inequality:} Let $X_1, X_2, \dots, X_n$ be independent random variables, each satisfying $|X_i - \mathbb{E}[X_i]| \leq B$. Bernstein’s inequality gives an upper bound on the probability that the sum of these variables deviates from its expected value by at least $t$, that is
\begin{equation}
    \Pr\left( \sum_{i=1}^n (X_i - \mathbb{E}[X_i]) \geq t \right) \leq \exp\left( \frac{-t^2}{2\left(\sum_{i=1}^n \sigma_i^2 + Bt/3\right)} \right),
\end{equation}
where $\sigma_i^2$ is the variance of each variable $X_i$.}

\medskip
\noindent
Compared to McDiarmid’s inequality, Bernstein’s inequality provides a tighter bound because it takes into account the variability of the data through the variance $\sigma_i^2$ of each random variable, leading to a more refined estimation of the p-value. 

\medskip
\noindent
However, in our protocol, we do not make assumptions about the specific quantum states involved. Consequently, we do not have direct access to the true variances $\sigma_i^2$ of the underlying variables. To address this challenge, we consider applying the Bathia-Davis inequality~\cite{Bhatia2000}, which provides an upper bound on the variance from known information about the mean and range of a random variable.

\medskip
\noindent
\textbf{Bathia-Davis inequality:} For any random variable $X$ with minimum value $a$, maximum value $b$, and mean $\mu = \mathbb{E}[X]$, the variance of $X$ satisfies 
\begin{equation}
\text{Var}(X) \leq (b - \mu)(\mu - a). 
\end{equation}

By using this inequality, we can estimate an upper bound for the variances of the random variables in our protocol. This allows us to proceed with Bernstein’s inequality even without direct knowledge of the variances, ensuring that we still obtain meaningful probabilistic bounds for the p-value in our analysis. To calculate the upper bound of the p-value $P(\widetilde{\Gamma}_{c} \leq \gamma_{c} | H_{0})$, we first state the following proposition. 

\begin{proposition}
\label{P1}
{Let $Z_1,...,Z_l$  be independent bounded stochastic variables, such that $\mathbb{E}(Z_i) = \mu_z$ and the realization satisfies $a \leq z_i\leq b $, with $a<0$ and $b>0$. Also let $Z = {\frac{1}{l}} \sum_{i=1}^l Z_i$ and $\sigma_{z}^{2} = {\frac{1}{l}}\sum_{i=1}^{l} \text{Var}(Z_i)$. Then for the realization $z={\frac{1}{l}} \sum_{i=1}^l z_i$ which satisfies $z<0$ and $\mu_z\geq 0$, one has}
\begin{equation}
    P(Z\leq z )\leq \exp{\dfrac{z^2 l}{2ba - \dfrac{2}{3}(b-a)z}}.
\end{equation}
\end{proposition}

\begin{proof}
Since $a\leq z_i \leq b$, we have that $|z_i - \mu_z|\leq b-a$. Then from Bernstein's inequality, we obtain the following one-sided inequality
\begin{equation}
    P(Z-\mu_z \leq -t)\leq  \exp{\dfrac{-t^2 l}{2\sigma_{z}^{2} + \dfrac{2}{3}(b-a)t}}, \text{with } t\geq0.
\end{equation}
The probability $P(Z\leq z )$, can then be calculated as
\begin{equation}
    P(Z\leq z )=P(Z-\mu_z\leq z-\mu_z )\leq \exp{-\dfrac{(z-\mu_z)^2 l}{2\sigma_{z}^{2} - \frac{2}{3}(b-a) (z-\mu_z) }}, \quad z<0\leq\mu_z.
\end{equation}
By using the Bathias-Davis inequality, we get
\begin{equation}
    \sigma_{z}^2 \leq \max_{i\in\left[1,...,l \right]}\text{Var}(z_i)\leq (b-\mu_z)(\mu_z -a). 
\end{equation}
Combining the above two inequalities, the probability $P(Z\leq z )$ can be bounded by
\begin{equation}
\label{ineq1}
    P(Z\leq z )\leq \exp{\dfrac{-(z-\mu_z)^2 l}{2(b-\mu_z)(\mu_z -a) - \dfrac{2}{3}(b-a)(z-\mu_z)}}.
\end{equation}
To analyze the monotonicity of the right-hand side of the above inequality with respect to $\mu_z\geq0$, we calculate the following derivative
\begin{equation}
   \frac{d}{d\mu_{z}} \dfrac{-(z-\mu_z)^2 }{2(b-\mu_z)(\mu_z -a) - \dfrac{2}{3}(b-a)(z-\mu_z)}=\frac{3 (\mu_z-z) [a (b-\mu_z-2z)+b (2a-2 \mu_z-z)+3 \mu_z z]}{\left[a (3 b-2 \mu_z-z)+b (z-4 \mu_z)+3\mu_z^2\right]^2}.
\end{equation}
Since $(b-\mu_z-2z)\geq0$ and $(2a-2 \mu_z-z)\leq0$, the derivative is less than or equal to 0. Therefore, the right-hand side of the  inequality~\ref{ineq1} is monotonic decreasing with $\mu_z$ when $\mu_z \geq 0$. The probability $P(Z\leq z )$ can then be further bounded by choosing $\mu_z=0$ as
\begin{equation}
    P(Z\leq z ) \leq \exp{\dfrac{z^2 l}{2ba + \dfrac{2}{3}(b-a)z}}.
\end{equation}  
\end{proof}

\medskip
\noindent
By using the Proposition~\ref{P1}, we can calculate the upper bound of the p-value $P(\widetilde{\Gamma}_{c} \leq \gamma_{c} | H_{0})$. First, based on the null assumption $H_{0}$, we have $\Gamma\geq0$. Since $\Gamma_c\geq\Gamma$, it follows that $\Gamma_{c}\geq0$, which satisfies the conditions of Proposition~\ref{P1}. 
Additionally, the values of $\Gamma_{0}^{c}$ given in the main text are always less than 0, and $\Gamma_{1}^{c}$ are always larger than 0, both of which satisfy the conditions for using Proposition~\ref{P1}. Therefore, $P(\widetilde{\Gamma}_{c} \leq \gamma_{c} | H_{0})$ can be upper bounded by

\begin{equation}
\label{upper bound h0}
P(\widetilde{\Gamma}_{c}\leq \gamma_{c} | H_0)\leq \exp{\dfrac{\gamma^{2}_{c} M/2}{2\Gamma^{c}_{1}\Gamma^{c}_{0}+2(\Gamma^{c}_{1}-\Gamma^{c}_{0})\gamma_{c}/3}}.
\end{equation}
Note that this bound does not depend on a specific quantum state, but applies to all states satisfying the condition $H_0$. Here, $\gamma_c$
serves as a variant of the original experimental observation
$\gamma$ and satisfies $\gamma\leq\gamma_c$ for any $c=(\alpha,\beta)$. Note that an important condition is that $\gamma_c<0$. In experiments, hypothesis testing is conducted only when $\gamma<0$ is observed. If certain values of $c=(\alpha,\beta)$ result in
$\gamma_c\geq0$, there is a risk of misclassifying a potentially spin-squeezed quantum state as not being spin-squeezed based on criterion $\Gamma_c$. Therefore, it is essential to set $\gamma_{c}<0$ to prevent such misclassification. This requirement also aligns with the operational criteria outlined in Proposition~\ref{P1}. In the following text, we assume that condition $\gamma_c<0$ is met by default.

\medskip
\noindent
Note that the bound \eqref{upper bound h0} is a consequence of $\Gamma_c\geq 0$. Therefore, the same bound also holds when considering the modified null hypothesis $H_{0}^{c}$ as ``the state of interest satisfies $\Gamma_{c}\geq0$", i.e.
\begin{equation}
\label{upper bound h0c}
P(\widetilde{\Gamma}_{c}\leq \gamma_{c} | H^{c}_0)\leq \exp{\dfrac{\gamma^{2}_{c} M/2}{2\Gamma^{c}_{1}\Gamma^{c}_{0}+2(\Gamma^{c}_{1}-\Gamma^{c}_{0})\gamma_{c}/3}}.
\end{equation}
This means that the same p-value applies to both hypotheses.

\section{Conservative assumptions for the estimations of the number of experimental repetitions}
In this section, we explain why the assumptions made during the estimation process are conservative, meaning they result in an underestimation of the number of measurements required to set an upper bound of 5\% on the p-value in previously conducted experiments. In our analysis, the criteria for testing spin squeezing is defined as 
\begin{equation}
\Gamma_{c}=
N\langle Q^{2}_{\vec{n}_{\perp}}\rangle -2N\alpha\langle Q_{\vec{n}_{\perp}}\rangle-2\beta\langle Q_{\vec{n}}\rangle+N\alpha^{2}+\beta^2. 
\end{equation}
Let the sample statistics be $S_{\Vec{n}_{\perp}}$, $\mu_{\Vec{n}_{\perp}}$ and $\mu_{\Vec{n}}$ for the quantities $\text{Var}(Q_{\vec{n}_{\perp}})$, $\langle Q_{\vec{n}_{\perp}}\rangle$ and $\langle Q_{\vec{n}}\rangle$. The estimation of $\Gamma_c$ can then be deduced from 
\begin{equation}
    \bar{\gamma}_{c}=N[S_{\Vec{n}_{\perp}}+{(\mu_{\Vec{n}_{\perp}})}^2]-2N\alpha\mu_{\Vec{n}_{\perp}}- 2\beta\mu_{\Vec{n}} +N\alpha^2+\beta^2.
\end{equation}
However, many references on experimental attempts to produce and detect spin squeezing only report the values of the variance $S_{\Vec{n}_{\perp}}$ and mean $\mu_{\Vec{n}}$, while the information on $\mu_{\Vec{n}_{\perp}}$ is either missing or not provided. Hence, we just compute the estimation of $\Gamma_c$ from 
\begin{equation}
\label{gammac}
    \gamma_c=N S_{\Vec{n}_{\perp}}-2\beta\mu_{\Vec{n}} +\beta^2.
\end{equation}

\medskip
\noindent
In theory, the value of $\mu_{\Vec{n}_{\perp}}$ should be zero if the measurements are performed in the ideal directions. However, due to experimental noise and statistical fluctuations, the observed value of $\mu_{\Vec{n}_{\perp}}$ may be slightly different from zero. Therefore, we consider all possible values in the range $\mu_{\Vec{n}_{\perp}}\in[-0.1,0.1]$. This represents a significant deviation given that $|\mu_{\Vec{n}_{\perp}}|\leq 1$. In practice, if a deviation $|\mu_{\Vec{n}_{\perp}}|= 0.1$ is observed, it may indicate that the chosen main direction is not appropriate.

\medskip
\noindent
For each dataset, the numerical calculations show that the required number of measurements increases as $|\mu_{\vec{n}_{\perp}}|$  increases.  The minimum number of measurements required to achieve a p-value of 0.05 occurs when $\mu_{\Vec{n}_{\perp}}=0$. Conversely, the maximum number of measurements is required when $|\mu_{\Vec{n}_{\perp}}|=0.1$. Therefore, setting $\mu_{\Vec{n}_{\perp}}=0$ offers a conservative estimate of the number of experimental repetitions required to place an upper bound of 5\% on the p-value.

\medskip
\noindent
The numbers of spins $N$ and experimental repetitions that we have considered are given in the Table~\ref{tb1} and~\ref{tb2} below. These tables list the number of measurement repetitions for each experiments~\cite{meyer2001,Bornet2023,Franke2023,Bohnet2016,Strobel2014,Riedel2010,Ockeloen2013,SchleierSmith2010,Leroux2010,LouchetChauvet2010,Bohnet2014,Sewell2012} (when not explicitly stated in the paper, we used the maximum number of measurements reported in the article), together with the number of experimental repetitions $M(\mu_{\Vec{n}_{\perp}}=0)$ $(M(\mu_{\Vec{n}_{\perp}}=0.1)$ that would be needed
to place an upper bound of 5\% on the p-value under the assumption that the mean $\mu_{\Vec{n}_{\perp}}=0$ ($\mu_{\Vec{n}_{\perp}}=0.1$).

% \begin{table}[ht]
%     \centering
%     \caption{The number of Measurements}
%     \label{tb1}
%     \begin{tabular}{cccccccccccc}
%         \toprule
%         System size & 2\cite{meyer2001} & 21\cite{Bohnet2016} & 58\cite{Bohnet2016} & 144\cite{Strobel2014} & 1250\cite{Riedel2010} & 1400\cite{Ockeloen2013} & 470\cite{Strobel2014} & 50000~\cite{Leroux2010} & 33000~\cite{SchleierSmith2010} & 90000~\cite{LouchetChauvet2010} & 740000~\cite{Sewell2012} \\
%         $M_0$ & 10000 & 400 & 400 & 400 & 740 & 240 & 32500  & 200  & 200 & 9600 & 2180 \\
%         $M (\mu_{\Vec{n}_{\perp}}=0.0)$  & 21200 & 8800 & 5340 & 13000 & 113800 & 145800 & 19970 & 1710400 & 2204400 & 8504400 & 118504400  \\
%         $M (\mu_{\Vec{n}_{\perp}}=0.1)$  & 23320 & 10460 & 6400 & 15600 & 137400 & 176400 & 24120 & 2070400 & 2670400 & 10270400 & 144070400 \\
%         \toprule
%     \end{tabular}
% \end{table}
% \begin{table}[ht]
%     \centering
%     \caption{The number of Measurements}
%     \label{tb2}
%     \begin{tabular}{ccccccccc}
%         \toprule
%         System size & 480000~\cite{Bohnet2014} & 12~\cite{Franke2023} & 4\cite{Bornet2023} & 9\cite{Bornet2023} & 16~\cite{Bornet2023} & 36~\cite{Bornet2023} & 64~\cite{Bornet2023} & 100~\cite{Bohnet2016}  \\
%         $M_0$ & 200 & 1200 & 400 & 400 & 400 & 400 & 400 & 400 \\
%         $M (\mu_{\Vec{n}_{\perp}}=0.0)$ & 21070400 & 728 & 3260 & 4160 & 5420 & 6760 & 10900 & 15600  \\
%         $M (\mu_{\Vec{n}_{\perp}}=0.1)$ & 25470400 & 842 & 3668 & 4840 & 6420 & 8080 & 13040 & 18800  \\
%         \toprule
%     \end{tabular}
% \end{table}
%aaaaaaaaaaaa

\begin{table}[ht]
    \centering
    \caption{The number of Measurements}
    \label{tb1}
    \begin{tabular}{cccccccccccc}
        \toprule
        System size & 2\cite{meyer2001} & 4\cite{Bornet2023} &  9\cite{Bornet2023} & 12~\cite{Franke2023} & 16~\cite{Bornet2023} & 21\cite{Bohnet2016} & 36~\cite{Bornet2023} & 58\cite{Bohnet2016} & 64~\cite{Bornet2023} & 100~\cite{Bohnet2016} & 144\cite{Strobel2014} \\
        Experimental measurements & 10000 & 400 & 400 & 1200 & 400 & 400 & 400  & 400  & 400 & 400 & 2180 \\
        $M (\mu_{\Vec{n}_{\perp}}=0.0)$  & 21200 & 3260 & 4160 & 728 & 5420 & 8800 & 6760 & 5340 & 10900 & 15600 & 13000  \\
        $M (\mu_{\Vec{n}_{\perp}}=0.1)$  & 23320 & 3668 & 4840 & 842 & 6420 & 10460 & 8080 & 6400 & 13040 & 18800 & 15600 \\
        \toprule
    \end{tabular}
\end{table}
\begin{table}[ht]
    \centering
    \caption{The number of Measurements}
    \label{tb2}
    \begin{tabular}{ccccccccc}
        \toprule
        System size & 470\cite{Strobel2014} & 1250\cite{Riedel2010} & 1400\cite{Ockeloen2013} & 33000~\cite{SchleierSmith2010} & 50000~\cite{Leroux2010} & 90000~\cite{LouchetChauvet2010} & 480000~\cite{Bohnet2014} & 740000~\cite{Sewell2012}  \\
        Experimental measurements & 32500 & 740 & 240 & 200 & 200 & 9600 & 200 & 2180 \\
        $M (\mu_{\Vec{n}_{\perp}}=0.0)$ & 19970 & 113800 & 145800 & 2204400 & 1710400 & 8504400 & 21070400 & 118504400  \\
        $M (\mu_{\Vec{n}_{\perp}}=0.1)$ & 24120 & 137400 & 176400 & 2670400 & 2070400 & 10270400 & 25470400 & 144070400  \\
        \toprule
    \end{tabular}
\end{table}

\section{Derivation of other upper bounds associated to $\Gamma$}
\label{other bounds}
In this section, we compare the upper bounds on the p-value that are obtained using various concentration inequalities. We demonstrate that, in comparison to using the original detection criterion $\Gamma$ (Eq.~3 of the main text), the family of criteria $\Gamma_c$ (Eq.~4 of the main text) provides a smaller upper bound on the p-value. This highlights the advantage of our approach in improving the statistical significance of experimental data.

\subsection{The criterion $\Gamma$ and McDiarmid upper bound}
\label{Mcbound}

We first continue the analysis started in Section~\ref{sec1} where we considered the criterion $\Gamma$. Specifically, we analyze the property of bounded differences in order to apply McDiarmid's inequality. This property is essential to establish concentration limits on the function $g$, which depends on independent random variables. %The calculation of the bounded differences is shown below. 

\medskip
\noindent
Let the realizations of the random variables $q_{{\Vec{n}}}^{(i)}$  and $q_{{\Vec{n}}_{\perp}}^{(j)}$ be $x_{i}$ and $y_{j}$, where $x_{i},y_{j}\in [-1,1]$, respectively. We thus have
\begin{align}
    &\sup_{x_i \in \mathcal{X}_i} \left|g(x_1, \dots, x_{k}, \dots, x_{M_{\vec{n}}}, y_{1}, \ldots, y_{M_{\vec{n}_{\perp}}}) - g(x_1, \dots, {x'}_{k}, \dots, x_{M_{\vec{n}}}, y_{1}, \ldots, y_{M_{\vec{n}_{\perp}}})\right| \\ \nonumber
    =&\sup_{x_i \in \mathcal{X}_i} \frac{N}{M_{\vec{n}}-1}\left[\frac{2}{M_{\vec{n}}}\sum_{i\neq k}{x_{i}}(x'_{k}-x_{k})+\frac{M_{\vec{n}}-1}{M_{\vec{n}}}(x_{k}^{2}-{x'}_{k}^{2})\right] \leq \frac{4N}{M_{\vec{n}}}=s_{x,k},
\end{align}
and
\begin{align}
    &\sup_{y_i \in \mathcal{Y}_i} \left|g(x_1, \dots, x_{M_{\vec{n}}}, y_{1}, \dots,y_{k},\dots, y_{M_{\vec{n}_{\perp}}}) - g(x_1, \dots, x_{M_{\vec{n}}}, y_{1}, \dots,{y'}_{k},\dots, y_{M_{\vec{n}_{\perp}}})\right| \\ \nonumber
    =&\sup_{y_i \in \mathcal{Y}_i} \frac{1}{M_{\vec{n}_{\perp}}-1}\left[\frac{2}{M_{\vec{n}_{\perp}}}\sum_{i\neq k}{y_{i}}(y'_{k}-y_{k})+\frac{M_{\vec{n}_{\perp}}-1}{M_{\vec{n}_{\perp}}}(y_{k}^{2}-{y'}_{k}^{2})\right]-\frac{1}{M_{\vec{n}_{\perp}}}(y_{k}^{2}-{y'}_{k}^2) \leq \frac{4}{M_{\vec{n}_{\perp}}}=s_{y,k}.
\end{align}
With the bounded difference $\{s_{x,k},s_{y,k}\}$, the significance level $P({\Gamma} \leq \gamma | H_{0})$ can be bounded by
\begin{equation}
    P(\widetilde{\Gamma} \leq \gamma | H_{0})= P(\widetilde{\Gamma}- \Gamma \leq \gamma-\Gamma | H_{0})\leq\exp{\frac{-2(\gamma-\Gamma)^{2}}{\frac{16N^{2}}{M_{\vec{n}}}+\frac{16}{M_{\vec{n}_{\perp}}}}}\leq \exp{\frac{-\gamma^{2}}{\frac{8N^{2}}{M_{\vec{n}}}+\frac{8}{M_{\vec{n}_{\perp}}}}}.
\end{equation}
The denominator in the above expression contains an $N^2$ term, which causes the p-value to increase rapidly as $N$ grows. This leads to a large upper bound on the p-value, necessitating more measurements to counterbalance the $N^2$ effect. The solution we propose in the main text is to focus on the family of criteria $\Gamma_c$, which is expressed as a sum of independent random variables and avoids the $N^2$ term introduced by the bounded differences. Below, we introduce an other approach based on Bernstein's inequality to deal with the $N^2$ problem.

\subsection{Bernstein's bound with $\Gamma$}

\begin{figure*}[ht]
    \centering
    \begin{minipage}{\textwidth} % Adjusting the width to full text width
        \centering
        \begin{tikzpicture}
            \node[anchor=south west,inner sep=0] (image) at (0,0) {\includegraphics[width=1\textwidth]{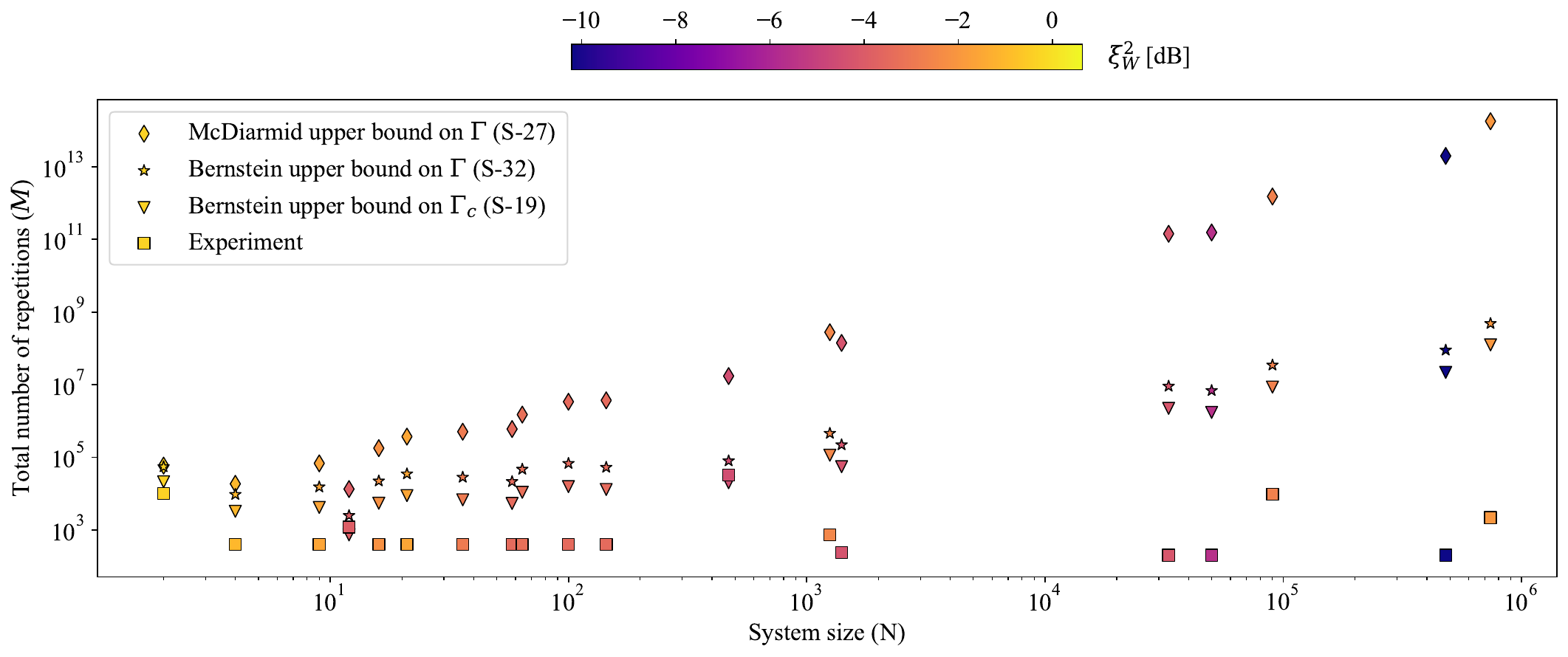}};
            \begin{scope}[x={(image.south east)},y={(image.north west)}]
                % Citations inside the plot
                \node[right] at (0.087,0.21) {\cite{meyer2001}};
                \node[right] at (0.12,0.15) {\cite{Bornet2023}};
                \node[right] at (0.176,0.15) {\cite{Bornet2023}};
                \node[right] at (0.207,0.16) {\cite{Franke2023}};
                \node[right] at (0.225,0.204) {\cite{Bornet2023}};
                \node[right] at (0.245,0.204) {\cite{Bohnet2016}};
                \node[right] at (0.28,0.204) {\cite{Bornet2023}};
                \node[right] at (0.304,0.20) {\cite{Bohnet2016}};
                \node[right] at (0.320,0.20) {\cite{Bornet2023}};
                \node[right] at (0.346,0.204) {\cite{Bornet2023}};
                \node[right] at (0.37,0.204) {\cite{Bohnet2016}};
                \node[right] at (0.45,0.23) {\cite{Strobel2014}};
                \node[right] at (0.51,0.223) {\cite{Riedel2010}};
                \node[right] at (0.533,0.165) {\cite{Ockeloen2013}};
                \node[right] at (0.724,0.20) {\cite{SchleierSmith2010}};
                \node[right] at (0.757,0.20) {\cite{Leroux2010}};
                \node[right] at (0.79,0.29) {\cite{LouchetChauvet2010}};
                \node[right] at (0.901,0.20) {\cite{Bohnet2014}};
                \node[right] at (0.93,0.25) {\cite{Sewell2012}};
            \end{scope}
        \end{tikzpicture}
    \end{minipage}
    % References below the plot in two columns using minipage
    \vspace{1em} % Some space before the references
    \begin{minipage}{0.32\textwidth}
        \begin{itemize}
            \item \cite{meyer2001} Meyer et al. (2001)
            \item \cite{Bornet2023} Bornet et al. (2023)
            \item \cite{Franke2023} Franke et al. (2023)
            \item \cite{Bohnet2016} Bohnet et al. (2016)
        \end{itemize}
    \end{minipage}
    \begin{minipage}{0.32\textwidth}
        \begin{itemize}
            
            \item \cite{Strobel2014} Strobel et al. (2014)
            \item \cite{Riedel2010} Riedel et al. (2010)
            \item \cite{Ockeloen2013} Ockeloen et al. (2013)
            \item \cite{SchleierSmith2010} Schleier-Smith et al. (2010)
        \end{itemize}
    \end{minipage}
    \begin{minipage}{0.32\textwidth}
        \begin{itemize}

            \item \cite{Leroux2010} Leroux et al. (2010)
            \item \cite{LouchetChauvet2010} Louchet-Chauvet et al. (2010)
            \item \cite{Bohnet2014} Bohnet et al. (2014)
            \item \cite{Sewell2012} Sewell et al. (2012)
        \end{itemize}
    \end{minipage}
 % Caption for the figure
    \caption{\textbf{Comparison of the number of experimental realizations based on different concentration inequalities.} The experiments reported in the literature are represented by squares, where the position indicates the total number of experimental repetitions (M) and the system size (N) that were realized. When calculating the number of measurements required to achieve a p-value of 0.05 using McDiarmid’s inequality (marked by diamonds), we assumed equal {number of} measurements in both measurement directions. The {stars} represent the Bernstein upper bound {applied to $\Gamma$}, calculated using S-32. The {downward triangles} correspond to the Bernstein upper bound, applied to the linearized parameter $\Gamma_c$, as reported presented in the main text.}
     \label{allinone}
\end{figure*}

We assume that 4 identical and independent experiments are performed, with half of them reporting on the measurement result of the angular momentum operator in the direction $\vec{n}$ while the rest is in the orthogonal direction $\vec{n_\perp}$. Then, from the collected data $\lbrace q_{\vec{n}}^{(1)},q_{\vec{n}}^{(2)}, q_{\vec{n_\perp}}^{(1)},q_{\vec{n_\perp}}^{(2)} \rbrace$, we estimate the variance using sample variance and the second moment by sample mean square. We get
\begin{equation}
   {\widetilde{ \Gamma'_1 }}= \dfrac{N}{2}\left( q_{\vec{n_\perp}}^{(1)}-q_{\vec{n_\perp}}^{(2)}\right)^2  - q_{\vec{n}}^{(1)}q_{\vec{n}}^{(2)}.
\end{equation}
This process is repeated $L=M/4$ times with $M$ the number of experimental realizations, such that the estimator of $\Gamma$ turns into the sum of random independent variables,
\begin{equation}
    \label{Bernstein estimator}
    {\widetilde{\Gamma'}} = \dfrac{1}{L}\sum_{i=1}^{L} {\widetilde{\Gamma'_i}},
\end{equation}
with 
\begin{equation}
    \label{berns estimator elements}
    {\widetilde{ \Gamma'_i }= \dfrac{N}{2}\left( q_{\vec{n_\perp}}^{(2i-1)}-q_{\vec{n_\perp}}^{(2i)}\right)^2  - q_{\vec{n}}^{(2i-1)}q_{\vec{n}}^{(2i)}.}
\end{equation}
It is easy to find that the estimator \eqref{Bernstein estimator} is unbiased:
 \begin{equation}
 \label{unbiased}
     \mathbb{E}({\widetilde{\Gamma'}}) = \Gamma.
 \end{equation}
Notice that the maximum and minimum values of \eqref{berns estimator elements} are $2N+1$ and $-1$, respectively. From the Proposition~\ref{P1}, we find
\begin{equation}
\label{bers_edi}
P({\widetilde{\Gamma'}}\leq \gamma | H_0)\leq \exp{-\dfrac{\gamma^2 M/8}{2N+1-  2\gamma (N+1)/3}}.
\end{equation}
From the previous result, the number of independent experimental realizations required to set an upper bound $p$ on the p-value is 
\begin{equation}
    M = 16\dfrac{\log{1/p}}{\gamma^2}\left[ \left(N+1\right)\left(1 - \dfrac{\gamma}{3}\right) - \dfrac{1}{2} \right].
\end{equation}
In Fig.~\ref{allinone}, we show the number of experiments based on different concentration inequalities. The method introduced in the main text {and detailed in Section \ref{sec:bestBound}, which relies on the linearized parameter $\Gamma_c$ together with Bernstein's bound, is the one that systematically} requires the fewest number of measurements. On the other hand, the approach based on {the parameter $\Gamma$ together with} McDiarmid's inequality requires the most measurements, as it employs the bounded differences property, which demands a large number of measurements to offset the $N^2$ effect. The Bernstein’s inequality to the parameter $\Gamma$ using  the technique where four independent measurements are considered in each round of the experiment {is also shown}. This allows 
$\Gamma$ to be expressed as a sum of independent random variables, and as a result, the p-value only contains the term 
$N$, without the $N^2$ component. Therefore, compared to the method based on McDiarmid’s inequality, the number of required measurements is reduced. This highlights the importance of a well-designed linear estimator, which can significantly reduce the required number of experimental measurements.

\section{Derivation of the lower bound on the p-value}
In this section, we focus on the lower bound of the p-value. We start with the mixed state
\begin{equation}
    \label{lower state}
    \hat{\rho} = r \ket{\chi}\bra{\chi} + \dfrac{1-r}{2}\left( \ket{\pmb{\uparrow_{\Vec{n}_{{\perp}}}}}\bra{\pmb{\uparrow_{\Vec{n}_{{\perp}}}}} + \ket{\pmb{\downarrow_{\Vec{n}_{{\perp}}}}}\bra{\pmb{\downarrow_{\Vec{n}_{{\perp}}}}}\right).
\end{equation}
Here $\ket{\chi}$ is a spin squeezed state with mean spin direction given by $\Vec{n}$. We also have $\ket{\pmb{\uparrow_{\Vec{n}_{{\perp}}}}}=\ket{\uparrow_{\Vec{n}_{{\perp}}}}^{\otimes N} $, $\ket{\pmb{\downarrow_{\Vec{n}_{{\perp}}}}}=\ket{\downarrow_{\Vec{n}_{{\perp}}}}^{\otimes N}$ and $0\leq r\leq 1$.  %A mix state $\ket{\pmb{\uparrow_{\Vec{n_{\perp}}}}}\bra{\pmb{\uparrow_{\Vec{n_{\perp}}}}} + \ket{\pmb{\downarrow_{\Vec{n_{\perp}}}}}\bra{\pmb{\downarrow_{\Vec{n_{\perp}}}}}$ is considered because it has a good performance that the expectations of $\rho$ only depends on $\ket{\chi}\bra{\chi}$,
We have
\begin{align}
    \langle Q_{\vec{n}} \rangle & = \text{Tr}[Q_{\vec{n}} \hat\rho] =r\langle Q_{\vec{n}} \rangle_{|\chi\rangle}, \\
    \langle Q_{\vec{n}_{\perp}} \rangle & = \text{Tr}[Q_{\vec{n}_{\perp}} \hat\rho] =0.
\end{align}
Since 
\begin{equation}
    Q_{\vec{n}_{\perp}} = \frac{1}{N}\sum_{i}^{N}\sigma_{\vec{n}_{\perp}}^{(i)},
\end{equation}
we have 
\begin{equation}
\bra{\pmb{\uparrow_{\Vec{n}_{{\perp}}}}}(Q_{\vec{n}_{\perp}})^{2}\ket{\pmb{\uparrow_{\Vec{n}_{{\perp}}}}}=\bra{\pmb{\downarrow_{\Vec{n}_{{\perp}}}}}(Q_{\vec{n}_{\perp}})^{2}\ket{\pmb{\downarrow_{\Vec{n}_{{\perp}}}}}=\frac{1}{N^{2}}\left(N + N^{2} -N \right)=1
\end{equation}
and 
\begin{equation}
    \text{Var}(Q_{\vec{n}_{\perp}})=r\langle Q_{\vec{n}_{\perp}}^2 \rangle_{|\chi\rangle} +1-r.
\end{equation}
With the above results, the Wineland parameter is given by
\begin{equation}
    \xi^2_W = \frac{rN\langle Q_{\vec{n}_{\perp}}^2 \rangle_{|\chi\rangle} +(1-r)N}{r^2\langle Q_{\vec{n}} \rangle_{|\chi\rangle}^2} =\frac{1}{r}\xi_W^2 \left(\ket{\chi} \right)+\frac{1-r}{r^2}\frac{N}{\langle Q_{\vec{n}} \rangle_{|\chi\rangle}^2}.
\end{equation}
Under the assumption that $\hat \rho$ is not a spin-squeezed state, $\xi^2_W$ should be  equal to or larger than 1. Therefore, the parameter $r$ for such a state satisfies 
\begin{equation}
    r^2 + \left(\frac{N}{\langle Q_{\vec{n}} \rangle_{|\chi\rangle}^2}-\xi_W^2 \left(\ket{\chi} \right)\right)r-\frac{N}{\langle Q_{\vec{n}} \rangle_{|\chi\rangle}^2}\leq0,
\end{equation}
which gives 
\begin{equation}
\label{rmax}
r \leq r_{\text{max}}= \dfrac{1}{2} \left( \kappa + \sqrt{\kappa^2 +\dfrac{4N}{\langle Q_{\Vec{n}}\rangle_{\ket{\chi}}^2} } \right)
\end{equation}
where $\kappa=\xi_W^2 \left(\ket{\chi} \right) - N /\langle Q_{\Vec{n}}\rangle_{\ket{\chi}}^2 $. {Note that $r_{\text{max}}$ can be lower bounded by
\begin{equation}
r_{\text{max}} \geq \underset{\kappa, \langle Q_{\Vec{n}}\rangle_{\ket{\chi}}^2}{\min} r_{\text{max}} = \frac{\sqrt{N^2+4N}-N}{2},
\end{equation}
achieved for $\xi_W^2=0$ and $\langle Q_{\Vec{n}}\rangle_{\ket{\chi}}^2=1$.}

\medskip
\noindent
It should {also} be noted that the state $\hat \rho$ may not be the best state to construct the lower bound, i.e. there could exist other quantum states that bring the lower bound of the p-value closer to the upper bound. Further exploration of alternative states could lead to a more accurate and tighter bound on the p-value.